\documentclass[fleqn,usenatbib]{mnras}

\usepackage{newtxtext,newtxmath}

\usepackage[T1]{fontenc}

\DeclareRobustCommand{\VAN}[3]{#2}
\let\VANthebibliography\thebibliography
\def\thebibliography{\DeclareRobustCommand{\VAN}[3]{##3}\VANthebibliography}

\usepackage{graphicx}	
	\graphicspath{{Figures/}}
	
\usepackage{amsmath}	

\usepackage{cancel}

\usepackage{ulem}
\usepackage{mathtools}
\usepackage{hyperref} 
	\hypersetup{colorlinks,citecolor=blue,linkcolor=blue,urlcolor=blue}
	\usepackage{etoolbox}
		\makeatother
	
\usepackage[capitalise]{cleveref}
	\crefname{equation}{Equation}{Equations}
	\crefname{figure}{Figure}{Figures}
	\crefname{table}{Table}{Tables}
	
	\newcommand{\crefalt}[1]{\namecref{#1}~\ref{#1}}
	
\usepackage{ulem}

\usepackage{etoolbox,siunitx}
    \robustify\bfseries
\usepackage{booktabs,times}
\usepackage{arydshln}
\usepackage{threeparttable}
\usepackage{xspace}

\usepackage{floatrow}
\usepackage[labelfont=bf,font=Large]{subfig}
\DeclareRobustCommand\rsout{%
  \bgroup
  \markoverwith{\textcolor{red}{\rule[0.5ex]{2pt}{0.4pt}}}%
  \ULon
}

\newcommand{\reflab}{{}_0}

\newcommand{\R}{R}							        
\newcommand{\Rref}{\R\reflab}						

\newcommand{\rhog}{\rho_{\mathrm{g}}}				
\newcommand{\rhod}{\rho_{\mathrm{d}}}				
\newcommand{\rhodj}{\rho_{\mathrm{d}j}}             

\newcommand{\deltav}{\Delta \textbf{v}}             
\newcommand{\deltavj}{\Delta \textbf{v}_{j}}        
\newcommand{\deltava}{\Delta \textbf{v}_{a}}
\newcommand{\deltavb}{\Delta \textbf{v}_{b}}
\newcommand{\deltavja}{\Delta \textbf{v}_{ja}}
\newcommand{\deltavka}{\Delta \textbf{v}_{ka}}
\newcommand{\deltavjb}{\Delta \textbf{v}_{jb}}

\newcommand{\deltavk}{\Delta \textbf{v}_{k}}
\newcommand{\deltavmat}{\Delta \!\textbf{V}}

\newcommand{\f}{\mathbf{f}}                         
\newcommand{\fg}{\mathbf{f}_{\mathrm{g}}}           
\newcommand{\fdj}{\mathbf{f}_{\mathrm{d}j}}         
\newcommand{\fdsum}{\mathbf{f}_{\mathrm{d}}}        
\newcommand{\fgvisc}{\mathbf{f}_{\mathrm{g,visc}}}

\newcommand{\deltafj}{\Delta \mathbf{f}_{j}}        

\newcommand{\vb}{\textbf{v}}  				        
\newcommand{\Vg}{v}						        	
\newcommand{\Vd}{v}						        	
\newcommand{\vg}{\textbf{\Vg}_{\mathrm{g}}}         
\newcommand{\vd}{\textbf{\Vd}_{\mathrm{d}}}         
\newcommand{\vdj}{\textbf{\Vd}_{\mathrm{d}j}}       
\newcommand{\vj}{\textbf{\Vd}_{j}}                  

\newcommand{\St}{\mathrm{St}}				    	
\newcommand{\Stj}{\mathrm{St}_j}					
\newcommand{\tj}{t_{j}}

\newcommand{\tja}{t_{ja}}
\newcommand{\tk}{t_{k}}

\newcommand{\tka}{t_{ka}}

\newcommand{\Kj}{K_{j}}
\newcommand{\Knj}{\mathrm{K}_{\mathrm{n}j}}         

\newcommand{\sj}{s_j}						    	
\newcommand{\smin}{s_{\mathrm{min}}}				
\newcommand{\smax}{s_{\mathrm{max}}}				

\newcommand{\epsj}{\epsilon_{j}}                    
\newcommand{\epsd}{\epsilon}
\newcommand{\epsda}{\epsilon_{a}}
\newcommand{\epsdja}{\epsilon_{ja}}

\newcommand{\epsdb}{\epsilon_{b}}
\newcommand{\epsdjb}{\epsilon_{jb}}

\newcommand{\vareps}{\varepsilon}					

\newcommand{\Sigmag}{\Sigma_{\mathrm{g}}}			
\newcommand{\Sigmagref}{\Sigmag\reflab}				

\newcommand{\cs}{c_{\mathrm{s}}}					
\newcommand{\csref}{\cs{}_{\mathrm{,}}\reflab}		

\newcommand{\Hg}{H}						           	

\newcommand{\Mstar}{M}						    	
\newcommand{\Mdisc}{M_{\mathrm{disc}}}		    	
\newcommand{\Mdj}{M_{\mathrm{d}j}}
\newcommand{\Mg}{M_{\mathrm{g}}}

\newcommand{\G}{\mathcal{G}}				    	
\newcommand{\alphaAV}{\alpha_{\mathrm{AV}}}	    	
\newcommand{\betaAV}{\beta_{\mathrm{AV}}}	    	

\newcommand{\sumj}{\sum_{j}}                        
\newcommand{\sumk}{\sum_{k}}                        

\newcommand{\s}{\; {\mathrm{s}}}					
\newcommand{\gram}{\; {\mathrm{g}}}			    	
\newcommand{\km}{\; {\mathrm{km}}}			    	
\newcommand{\cm}{\; {\mathrm{cm}}}			    	
\newcommand{\mm}{\; {\mathrm{mm}}}			    	
\newcommand{\mum}{\; \mu {\mathrm{m}}}		    	
\newcommand{\au}{\; {\mathrm{au}}}					
\newcommand{\yr}{\; {\mathrm{yr}}}					

\newcommand{\percent}{\% }

\newcommand{\matr}[1]{\mathbf{#1}}

\newcommand{\Phantom}{\textsc{Phantom}\xspace}

\defcitealias{LP12a}{LP12a}
\defcitealias{LP12b}{LP12b}
\defcitealias{LP14a}{Paper~I}
\defcitealias{NSH86}{NSH86}

\setcitestyle{notesep={; }}

\title[Full one-fluid multigrain]{Full one-fluid dusty gas with multiple grain species in SPH}

\author[Hutchison, Laibe, et al.]{
Mark Hutchison,$^{1,2}$\thanks{E-mail: markahutch@gmail.com}
Guillaume Laibe,$^{3}$
Giovanni Tedeschi-Prades,$^{2}$
Timoth{\'e}e David-Cl{\'e}ris,$^{4}$
\newauthor
Alex Barret,$^{3}$
Maxime Lombart,$^{5}$
Daniel J. Price,$^{6}$
and Christine M. Koepferl$^{1,2}$
\\
$^{1}$Munich University of Applied Sciences HM, Department for Technical Systems, Processes and Communication, Lothstra{\ss}e 34, 80335 M{\"u}nchen, Germany \\
$^{2}$Universit{\"a}ts-Sternwarte, Ludwig-Maximilians-Universit{\"a}t  M{\"u}nchen, Scheinerstr. 1, 81679 M{\"u}nchen, Germany \\
$^{3}$Univ Lyon, Univ Lyon1, Ens de Lyon, CNRS, Centre de Recherche Astrophysique de Lyon UMR5574, F-69230, Saint-Genis-Laval, France\\
$^{4}$CNRS, IPAG, Université Grenoble Alpes, F-38000 Grenoble, France\\
$^{5}$Universit{\'e} Paris-Saclay, Universit{\'e} Paris Cit{\'e}, CEA, CNRS, AIM, F-91191 Gif-sur-Yvette, France\\
$^{6}$School of Physics and Astronomy, Monash University, Clayton VIC 3800, Australia
}

\date{Accepted XXX. Received YYY; in original form ZZZ}

\pubyear{\the\year{}}

\begin{document}
\label{firstpage}
\pagerange{\pageref{firstpage}--\pageref{lastpage}}
\maketitle

\begin{abstract}
We present a Smoothed Particle Hydrodynamics (SPH) implementation of the full one-fluid dusty gas algorithm for multiple dust species, generalising our previous terminal velocity approach to handle arbitrary drag regimes. By construction, mass, momentum, angular momentum, and energy are all conserved. We benchmark our method against a suite of tests -- \textsc{dustybox}, \textsc{dustywave}, \textsc{dustyshock}, \textsc{dustysettle}, and \textsc{dustydisc} -- each probing different aspects of the algorithm. Compared to the terminal velocity approximation, the full one-fluid approach incurs a computational cost increase of a factor of five to ten due to the added overhead of evolving the differential velocities and solving the drag terms implicitly. However, it accurately recovers analytic behaviour in regimes where the terminal velocity approximation fails. In such cases, errors from the terminal velocity approximation accumulate and propagate to other dust phases. We show that the stopping-time limiter commonly used in the terminal velocity approximation for numerical stability can substantially affect simulations containing large grains (Stokes numbers $\gtrsim 1$). While disabling the limiter leads to different outcomes, the discrepancy with the full one-fluid solution remains comparable, underscoring the importance of using a more general formulation for large grains. The full one-fluid formalism may be useful when including processes such as coagulation and fragmentation, where accurate treatment of large grains becomes essential. While the inability to model orbit-crossing dust trajectories remains a key limitation of the one-fluid formalism, this may eventually be addressed through the introduction of an effective dust pressure, mirroring how fluid models encapsulate microscopic velocity dispersion in gases.
\end{abstract}




\section{Introduction}
\label{sec:introduction}

`Dust' in astrophysical contexts is a loose term that refers to particles or solids with dimensions that vary from nm-sizes (e.g. aerosols, hazes, clouds) to m-sizes (e.g. pebbles, boulders, meteorites). With such a broad definition, it is unsurprising that at least some form of dust is present at almost all scales of the universe (e.g. planetary, interstellar, intergalactic). But more than being a benign bystander, the presence or absence of dust can, in many cases, influence the evolutionary outcome of large-scale phenomenon. For example, dust alters the chemical and thermal budget in planetary atmospheres \citep[e.g.][]{Gierasch/Goody/1972,Guha/etal/2021,Mehta/etal/2024,Haberle/etal/2025}, circumstellar and AGN accretion discs \citep[e.g.][]{Savvidou/Bitsch/Lambrechts/2020,Gavino/etal/2023,Soliman/Hopkins/2023}, and giant molecular clouds \citep[e.g.][]{Soliman/Hopkins/Grudic/2024}; drives stellar winds from AGB stars to galactic outflows \citep[e.g.][]{Thompson/etal/2015,Arakawa/etal/2022,Nakazato/Ferrara/2024}; facilitates molecule formation in the ISM \citep[e.g.][]{Minissale/etal/2016,Ceccarelli/etal/2023}; and provides the building blocks to form planets \citep[see][and references therein]{Liu/Ji/2020}. Moreover, depending on the context, dust can either aid or impede observations of distant objects. On one hand, emission from warm dust grains is a key observable in protoplanetary discs and star-forming regions \citep[e.g.][]{Andrews/etal/2018}, including indirect evidence of magnetic fields \citep[e.g.][]{Pelgrims/etal/2024,Ohashi/etal/2025}; on the other hand, interstellar/galactic dust is a major source of extinction and reddening of stars and galaxies \citep[e.g.][]{Riguccini/etal/2011,Koepferl/etal/2017, Beuther/etal/2023}. Thus, quantifying the size distribution and spatial concentration of potential dust grains remains an ongoing quest for many areas of astrophysics.  

The basic theory of how dust grains dynamically interact with a gas has long been known \citep{Stokes/1851,Epstein/1924}, but typically require hydrodynamical simulations to accurately predict the long-term dust evolution -- especially in turbulent environments \citep[e.g.][]{Stepinski/Valageas/1996,Tricco/Price/Laibe/2017}. Differential velocities between individual dust grains resulting from their unique drag histories with the gas lead to collisional growth \citep[e.g.][]{Smoluchowski/1916} or fragmentation \citep[e.g.][]{Safronov/1972,Cheng/Redner/1990,Kostoglou/Karabelas/2000} -- both well studied fields in their own right. As collisions can occur between dust grains of any size (including their own), the size evolution of all dust species are interconnected. While the gas strongly influences the collision velocities between grains, the backreaction from the changing dust distribution alters the dynamics of the gas. In turn, the gas communicates those dynamical changes back onto other dust grains, indirectly coupling the dynamics of all dust species \citep[an effect exemplified by the polydisperse streaming instability; see e.g.][]{Krapp/etal/2019,Paardekooper/McNally/Lovascio/2020}. Accounting for the gas-dust interactions from \textit{all} dust grains in numerical simulations is therefore important in capturing the full gamut of physical outcomes that develop out of this non-linear spiderweb of dependencies.

Building on this understanding, some 3D hydrodynamical simulations have already begun to directly couple dust growth with dust–gas dynamics \citep{Bate/Hutchison/Elsender/2026,Lombart/Lebreuilly/Maury/2026}. However, accurately resolving grain growth requires a large number of size bins and incurs substantial computational cost, making it difficult to simulate objects on long timescales. Furthermore, extending models to larger grains sizes with more realistic coagulation-fragmentation physics still remains a fundamental challenge. Thus, before tackling the complex problem of coupling coagulation and fragmentation with 3D hydrodynamics, we first need efficient algorithms to model these processes independently.

Significant progress has already been made in developing accurate, efficient, and modular coagulation and fragmentation solvers \citep{Lombart/Laibe/2021, Lombart/Hutchison/Lee/2022, Lombart/etal/2024}, which are strong candidates for integration with hydrodynamic codes. Equally important is the development of a hydrodynamic framework that can (i) capture the full range of grain dynamics produced by these solvers and (ii) facilitate mass transfer between size bins with minimal computational overhead.
Using the reference frame of the gas-dust barycentre, \citet{Laibe/Price/2014c} proposed a `multigrain' one-fluid framework for modelling multiple dust phases in a gas on the same set of simulation particles without restrictions on the Stokes number.

The lack of restriction on drag regime is sometimes referred to as the `full' model to distinguish it from later variations that rely on the `terminal velocity approximation' \citep[e.g.][]{Youdin/Goodman/2005}. For example, \citet{Hutchison/Price/Laibe/2018} implemented a multigrain one-fluid model using the terminal velocity approximation in the SPH code \Phantom \citep{Price/etal/2018}, generalizing the earlier single-species version from \citet{Price/Laibe/2015}. It is this basic algorithm that was employed in \citet{Bate/Hutchison/Elsender/2026} for the dust dynamics, together with a new implicit solver for the diffusion equation \citep{Elsender/Bate/2024} that resolves several numerical limitations of the original implementation, including the removal of the drag timestep constraint, the reliance on positivity-enforcing formulations, and the need for a stopping-time limiter \citep[see][]{Ballabio/etal/2018}. Since then, similar multi-dust schemes have been implemented in other codes \citep[e.g.][]{Lebreuilly/Commercon/Laibe/2019,Lovascio/Paardekooper/2019}. However, one of the drawbacks of these implementations is that the terminal velocity approximation starts exhibiting errors on the order of 10\percent for grains with Stokes numbers of $\St \sim 0.1$ and 100\percent for $\St \gtrsim 1$ \citep{Laibe/Price/2014a}. In protoplanetary discs,\footnote{We will focus our discussion around protoplanetary discs for the remainder of this work, but similar scenarios/arguments can be found for the many other astrophysical environments mentioned earlier.} grains in this regime are routinely created by coagulation processes \citep[e.g.][]{Birnstiel/Dullemond/Brauer/2010}, but also quickly removed due to drift \citep[e.g.][]{Weidenschilling/1977}. Since it is usually assumed that much of the dust mass is locked away in larger grain sizes \citep[e.g.][]{Mathis/Rumpl/Nordsieck/1977}, an accurate accounting of the dynamics of large grains could be vitally important -- especially if this mass is later released through fragmentation at a different location. Thus, while much can still be achieved using the terminal velocity approximation, moving beyond this approximation will be required to capture the full physics in next-generation codes that couple hydrodynamics with dust coagulation and fragmentation.

Some codes are already capable of modelling multiple dust species spanning all drag regimes \citep{Benitez-Llambay/Krapp/Pessah/2019,Krapp/Benitez-Llambay/2020,Stoyanovskaya/etal/2021,Huang/Bai/2022,Krapp/etal/2024,Verrier/Lebreuilly/Hennebelle/2025,Sewanou/Laibe/Commercon/2025}, including \Phantom \citep{Mentiplay/etal/2020}, the code we are using for this study. In this last instance, the gas and any small dust species are modelled together on one set of simulation particles \citep[i.e.][]{Hutchison/Price/Laibe/2018} and each additional large dust species as a separate set of simulation particles \citep[based on the work of][]{Laibe/Price/2012a,Laibe/Price/2012b,Price/Laibe/2020}. Utilising different algorithms to treat the small and large dust grains separately circumvents the numerical challenges that arise at the limits of either method, namely: simulating tightly-coupled grains using separate particles from the gas (see discussion in \citealt{Laibe/Price/2012a,Laibe/Price/2012b}; in large part solved by \citealt{Price/Laibe/2020}) and loosely-coupled grains using the same particles as the gas (see discussion in \citealt{Laibe/Price/2014b}; although see \citealt{Stoyanovskaya/etal/2021} for a potential solution on how one can handle locally multi-valued velocities). However, in the context of coagulation and fragmentation, the multigrain formalism is generally better suited to handle collisions (with the exception of interpenetrating dust flows) and mass transfer between size bins than the multiple dust-particle approach because (i) the differential velocities between dust phases in the multigrain method are trivially calculated without the need for interpolation (vital information for determining the collisional outcomes of colliding dust grains), (ii) it does not require altering the mass or number of simulation particles and (iii) it maintains equal spatial resolution for all dust sizes, regardless of the total mass in those size bins \citep[vital when trace species are observationally important, such as $\mum$-sized presolar grains with strong nucleosynthetic signatures; see][]{Hutchison/etal/2022}. For these reasons, in the present work we return to the original \citet{Laibe/Price/2014c} model to derive and implement the corresponding SPH relations in \Phantom, benchmarking our work using standard dust tests.

The structure of the paper is as follows. In \cref{sec:evoleqone} we provide an overview of the relevant one-fluid variables and fluid equations. In \cref{sec:SPH} we summarise the final one-fluid SPH equations derived in \cref{sec:sph_derivation} and later implemented into \Phantom, including a description of our implicit drag solver. \Cref{sec:benchmarking_tests} contains the results from our benchmarking tests and \cref{sec:conclusions} our conclusions. Finally, \cref{sec:dustywave_solution} provides details of the 1D numerical solver we used to validate our \textsc{dustywave} results.

\section{Multigrain formalism}
\label{sec:evoleqone}

In this section, we introduce the relevant physical quantities and relations from \citet{Laibe/Price/2014c} that we will later need to derive the SPH relations. In everything that follows, we consider a system that consists of a mixture of a single gas phase and $N$ dust phases. Subscripts $\mathrm{g}$ and $\mathrm{d}$ distinguish between gas and dust properties, respectively, while subscripts $j$ and $k$ identify the relevant dust phase.

\subsection{Physical quantities}
\label{sec:phys_quant}

The total density of a mixture of fluids, $\rho$, is defined as the sum of densities of its constituent parts (i.e. gas and dust),
\begin{equation}
    \rho \equiv \rhog + \sumj \rhodj = \rhog + \rhod   ,
\end{equation}
and is advected at the barycentric velocity of the mixture,
\begin{equation}
    \vb \equiv \frac{\rhog \vg + \sumj \rhodj \vdj }{\rho} .
    \label{eq:def_vb}
\end{equation}
We can quantify the relative abundance of each dust phase by defining a dust fraction\footnote{Not to be confused with the dust-to-gas ratio, which is also frequently represented in the literature using the Greek letter $\epsilon$ or $\vareps$.} for each species,
\begin{equation}
    \epsj \equiv \frac{\rhodj}{\rho} ,
\end{equation}
such that their sum or total dust fraction,
\begin{equation}
    \epsd \equiv \sumj \epsj = \frac{\rhod}{\rho} ,
    \label{eq:def_epsilon}
\end{equation}
gives the relative dust mass to total mass of the mixture (the corresponding gas fraction being $1 - \epsd$).
Note that by adopting the definitions above, the total dust velocity, $\vd$, for arbitrary $N$ should be interpreted as the mass weighted sum of the individual dust velocities
\begin{equation}
    \vd = \frac{1}{\epsd} \sumj \epsj \vj .
\end{equation}
The same can be done for the differential velocities between the gas and individual dust phases $\deltavj \equiv \vdj - \vg$, namely
\begin{equation}
    \deltav = \frac{1}{\epsd} \sumj \epsj \deltavj .
    \label{eq:def_deltav}
\end{equation}
The above definitions allow us to rewrite \cref{eq:def_vb} more compactly as
\begin{equation}
	\vb =  (1-\epsd) \vg + \epsd \vd ,
\end{equation}
or, vice versa, the gas and dust velocities in terms of barycentric quantities
\begin{align}
    \vg & =  \vb - \epsd \deltav ,
    \label{eq:vgas}
\\
    \vdj & = \vb - \epsd \deltav + \deltavj .
    \label{eq:vdustj}
\end{align}
Taking $u$ to be the specific thermal energy of the gas, we can then write the specific energy density $e$ of the mixture as
\begin{equation}
    e =  \frac{1}{2} \vb^{2} + \frac{1}{2} \left[ \sumj \epsj \deltavj^{2} - \left( \epsd \deltav \right)^{2} \right] + \left(1 - \epsd \right) u.
\end{equation}
%

\subsection{Continuum equations}
\label{sec:eq_of_evol}


Using the physical quantities outlined above, \citet{Laibe/Price/2014c} showed that the fluid equations for a gas-dust mixture can be written in the barycentric reference frame as follows\footnote{The original \citet{Laibe/Price/2014c} paper contained two errors in the continuum relations for the evolution of the barycentric and differential velocities. In \cref{eq:genmomentum_bary} we have made a minor adjustment in vector arrangement to correct for the fact that $\deltav \deltavj \ne \deltavj \deltav$. More importantly, we have added missing terms to \cref{eq:genmomentum_deltav} that were absent in the original paper. Note the equivalent terms in the \citet{Laibe/Price/2014b} paper for a single species were also missing, which were later corrected in \cite{Lebreuilly/Commercon/Laibe/2019}.}
\begingroup
\allowdisplaybreaks
\begin{align}
    \frac{{\rm d} \rho}{{\rm d} t} = {}&  - \rho (\nabla \cdot \vb)  , 
    \label{eq:genmass_rho}
\\
    \frac{{\rm d} \epsj}{{\rm d} t} = {}&  -\frac{1}{\rho} \nabla \cdot \left[ \rho\epsj \left( \deltavj - \epsd \deltav \right) \right]
    \label{eq:gendtgevol} , 
\\
    \frac{{\rm d} \vb}{{\rm d} t}  = {}&  \left(1 - \epsd \right)\fg + \sumj \epsj \fdj + \f
    \nonumber
\\
    {}& - \frac{1}{\rho}\nabla\cdot \Big[ \rho \sumj  \epsj \left(\deltavj - \epsd \deltav \right) \deltavj  \Big] 
    \label{eq:genmomentum_bary},
\\
    \frac{{\rm d} \deltavj}{{\rm d} t}  = {}&{~} 
    \deltafj
    - (\deltavj \cdot \nabla) \vb
    - \frac{1}{2}\nabla \left[ \deltavj \cdot \left(\deltavj - 2 \epsd \deltav \right) \right]
    \nonumber
\\
    {}& 
    - \nabla (\deltavj) \cdot (\epsd \deltav)
    + \nabla (\deltavj - \epsd \deltav) \cdot \deltavj
    + (\epsd \deltav \cdot \nabla) \deltavj
    \nonumber
\\
    {}&  
    - (\deltavj \cdot \nabla) (\deltavj - \epsd \deltav)
    - \frac{\deltavj}{\epsj \tj} -  \sum_{k} \frac{\deltavk}{\left(1 - \epsd \right)\tk} ,
    \label{eq:genmomentum_deltav}
\\
    \frac{{\rm d} u}{{\rm d} t}  = {}&   -\frac{ P}{\rhog} (\nabla \cdot \vg) +  (\epsd \deltav \cdot \nabla) u 
     + \sumj   \frac{\deltavj^{2}}{\left(1 - \epsd \right)\tj}  .
    \label{eq:newusingle}
\end{align}
\endgroup
Here the comoving derivative refers to a particle moving with the barycentric velocity $\vb$, i.e.
\begin{equation}
    \frac{{\rm d}}{{\rm d}t} \equiv \frac{\partial}{\partial t} + (\vb \cdot \nabla) ,
\end{equation}
$\f$ represents accelerations acting on both components of the fluid ($\fg$ and $\fdj$ for accelerations acting on the gas and dust components, respectively), $\deltafj \equiv \fdj - \fg$ is the differential force between the gas and each dust phase, $P$ is the gas pressure, and $\tj$ is a drag time-scale specific to each grain type,
\begin{equation}
    \tj \equiv \frac{\rho}{\Kj} .
    \label{eq:dragtime}
\end{equation}
In general, the drag coefficient $\Kj$ between the gas and each dust phase depends on local properties of the gas and dust, as well as the unique properties of each dust grain. Unless otherwise noted, we follow \citet{Stepinski/Valageas/1996} and determine the drag regime based on the Knudsen number of each species,
\begin{equation}
    \Knj = \frac{9 \lambda_{\mathrm{g}}}{4 \sj},
\end{equation}
where, assuming the grains are spherical, $\sj$ is the radius of grains in the $j$th bin and $\lambda_{\mathrm{g}}$ is the mean free path of the gas. Grains for which $\Knj \ge 1$ are in the Epstein regime, while grains with $\Knj < 1$ are in the Stokes regime. Details about each drag regime, including the exact form $\Kj$ takes in \Phantom, can be found in Sections 2.13.6 and 2.13.7 (and references therein) of \citet{Price/etal/2018}.

\section{SPH formalism}
\label{sec:SPH}

Each SPH particle contains a mixture of gas and dust whose relative concentrations vary in time but whose total mass, $m$, remains fixed. Particles are advected at the barycentric velocity of their components (i.e. Equation~\ref{eq:def_vb}), so a particle with position vector ${\bf x}$ would by definition obey $\frac{{\rm d}{\bf x}_{a}}{{\rm d}t} = {\bf v}_{a}.$ Here and throughout the paper we refer to individual SPH particles using indices $a$ and $b$.

Deriving the SPH relations that conserve total mass, gas mass, dust mass of each species, total linear momentum, and total energy while solving \cref{eq:genmass_rho,eq:gendtgevol,eq:genmomentum_bary,eq:genmomentum_deltav,eq:newusingle} is a necessary, but tedious process through which not all readers may wish to wade. Therefore, here we list only the final SPH relations and refer interested readers to \cref{sec:sph_derivation} for the full derivation
\begingroup
\allowdisplaybreaks
\begin{align}
    & \rho_{a} =  \sum_{b} m_{b} W_{ab} (h_{a}),
    \label{eq:final_rho}
\\
    & \frac{{\rm d}\epsdja}{{\rm d} t}  =  - \sum_{b} m_{b} \left[ \frac{\epsdja }{\Omega_{a} \rho_{a}} \left(\deltavja - \epsda \deltava \right)\cdot\nabla_{a} W_{ab} (h_{a})\right.
    \nonumber
    \\
    & \phantom{\frac{{\rm d}\epsdja}{{\rm d} t}  =- \sum_{b} m_{b}}
    + \left.  \frac{\epsdjb }{\Omega_{b} \rho_{b}}\left(\deltavjb - \epsdb \deltavb \right)\cdot \nabla_{a} W_{ab} (h_{b}) \right],
    \label{eq:final_dustfrac}
\\
    & \frac{{\rm d} \vb_{a}}{{\rm d} t} =  (1-\epsda) \fg + {\bf f}_{a}
    \nonumber
    \\
    & \phantom{\frac{{\rm d} \vb_{a}}{{\rm d} t} = {}} -\sum_{b} m_{b} \sumj \left[ \frac{ \epsdja \deltavja}{\Omega_{a} \rho_{a}} \left(\deltavja - \epsda \deltava \right) \cdot \nabla_{a} W_{ab}(h_{a}) \right.
    \nonumber
    \\
    &  \phantom{\frac{{\rm d} \vb_{a}}{{\rm d} t} =  -\sum_{b} m_{b} } + \left.\frac{\epsdjb\deltavjb}{\Omega_{b} \rho_{b}} \left(\deltavjb - \epsdb \deltavb \right) \cdot \nabla_{a} W_{ab}(h_{b}) \right] ,
    \label{eq:final_momentum}
\\
    & \frac{{\rm d}\deltavja}{{\rm d}t} = - \frac{\deltavj}{\epsj \tj} -  \sum_{k} \frac{\deltavk}{\left(1 - \epsd \right)\tk} - \fg
    \nonumber
    \\
    & \phantom{\frac{{\rm d}\deltavja}{{\rm d}t} = } + \frac{1}{\rho_{a}\Omega_{a}}\sum_{b} m_{b} {\bf v}_{ab}  \deltavja \cdot \nabla_{a} W_{ab} (h_{a})
    \nonumber
    \\
    & \phantom{\frac{{\rm d}\deltavja}{{\rm d}t} = } + \frac{1}{2\rho_{a}\Omega_{a}}  \sum_{b} m_{b}  \Big[ \deltavja \cdot \left(\deltavja - 2 \epsda \deltava \right) 
    \nonumber
    \\
    &  \phantom{\frac{{\rm d}\deltavja}{{\rm d}t} = + \frac{1}{2\rho_{a}\Omega_{a}}} -  \deltavjb \cdot \left(\deltavjb - 2 \epsdb \deltavb \right)  \Big] \nabla_{a} W_{ab} (h_{a})
    \nonumber
    \\
    & \phantom{\frac{{\rm d}\deltavja}{{\rm d}t} = } + \frac{1}{\rho_{a} \Omega_{a}} \sum_{b} m_{b} \epsda \deltava \cdot (\deltavja - \deltavjb) \nabla_{a} W_{ab} (h_{a})
    \nonumber
    \\
    & \phantom{\frac{{\rm d}\deltavja}{{\rm d}t} = }  - \frac{1}{\rho_{a} \Omega_{a}} \sum_{b} m_{b} \deltavja \cdot \Big[(\deltavja - \epsda \deltava) 
    \nonumber
    \\
    & \phantom{\frac{{\rm d}\deltavja}{{\rm d}t} =  - \frac{1}{\rho_{a} \Omega_{a}} \sum_{b} m_{b} }
    - (\deltavjb - \epsdb \deltavb) \Big] \nabla_{a} W_{ab} (h_{a}) 
    \nonumber
    \\
    &\phantom{\frac{{\rm d}\deltavja}{{\rm d}t} = } - \frac{1}{\rho_{a} \Omega_{a}} \sum_{b} m_{b} (\deltavja - \deltavjb) \epsda \deltava \cdot \nabla_{a} W_{ab} (h_{a}) 
    \nonumber
    \\
    & \phantom{\frac{{\rm d}\deltavja}{{\rm d}t} = } + \frac{1}{\rho_{a} \Omega_{a}} \sum_{b} m_{b} \Big[(\deltavja - \epsda \deltava)
    \nonumber
    \\
    & \phantom{\frac{{\rm d}\deltavja}{{\rm d}t} = - \frac{1}{\rho_{a} \Omega_{a}}  }
    - (\deltavjb - \epsdb \deltavb) \Big] \deltavja \cdot  \nabla_{a} W_{ab} (h_{a}) ,
    \label{eq:final_deltav}
\\
    & \frac{{\rm d}u_{a}}{{\rm d}t} = \sumj \frac{\deltavja^{2}}{\left(1 - \epsda \right)\tja} +\frac{P_{a} + q^{\rm AV}_{v, a}}{\Omega_{a} \rho_{a} \rho_{{\rm g},a}} \sum_{b} m_{b} {\bf v}_{{\rm g},ab} \cdot \nabla_{a} W_{ab}(h_{a})
    \nonumber
    \\
    & \phantom{\frac{{\rm d}u_{a}}{{\rm d}t} =} - \frac{1}{\Omega_{a} \rho_{a}}  \sum_{b} m_{b} u_{ab} \epsda \deltava \cdot \nabla_{a} W_{ab} (h_{a})
    \nonumber
    \\
    & \phantom{\frac{{\rm d}u_{a}}{{\rm d}t} =} + \frac{1}{1-\epsilon_{a}} \sum_{b} m_{b}
		\left[ \frac{Q^{\rm AC}_{u, a}}{\Omega_{a} \rho_{a}^{2}} F_{ab} (h_{a}) + 
		\frac{Q^{\rm AC}_{u, b}}{\Omega_{b} \rho_{b}^{2}} F_{ab} (h_{b}) \right],
  \label{eq:final_energy}
\end{align}
\endgroup
where
\begin{align}
    (1-\epsda) \fg = -\sum_{b} m_{b} &\left[ \frac{P_{a} + q^{\rm AV}_{v, a}}{\Omega_{a} \rho_{a}^{2}} \nabla_{a} W_{ab}(h_{a}) \right.
    \nonumber
\\
    & \qquad \qquad \left.
    + \frac{P_{b} + q^{\rm AV}_{v, b}}{\Omega_{b} \rho_{b}^{2}} \nabla_{a} W_{ab}(h_{b}) \right] ,
\end{align}
$W_{ab}(h) \equiv W(|\mathbf{r}_a- \mathbf{r}_b|,h)$ is the SPH kernel, $h$ is the smoothing length, and $\Omega$ is the usual term to account for smoothing length gradients
\begin{equation}
	\Omega_{a} = 1 - \frac{\partial h_{a}}{\partial \rho_{a}} \sum_{b} m_{b} \frac{\partial W_{ab} (h_{a})}{\partial h_{a}}.
\end{equation}
The artificial viscosity, differential velocity dissipation, and conductivity parameters are given by
\begin{align}
    q^{\rm AV}_{v, a} &\equiv
	\begin{cases}
		-\frac{1}{2} \left( 1 - \epsilon_{a} \right) v^{\mathrm{sig}}_{v,a} {\bf v}_{ab}^{\rm g} 
			\cdot \hat{{\bf r}}_{ab}, & \qquad {\bf v}_{ab}^{\rm g} \cdot \hat{{\bf r}}_{ab} < 0
	\\	
		0,  & \qquad \mathrm{otherwise},
	\end{cases}
    \label{eq:qav_v}
\\
    Q^{\rm AC}_{u,a} &\equiv \frac{1}{2} \alpha_u \rho_{a} v^{\mathrm{sig}}_{u} \left( u_{a} - u_{b} \right),
    \label{eq:qac_u}
\end{align}
with their corresponding signal speeds defined as
\begin{align}
    v^{\mathrm{sig}}_{v,a} &\equiv \alpha_{v,a} c_{\mathrm{s},a} + \beta_{v} \vert {\bf v}_{ab}^{\rm g} \cdot \hat{{\bf r}}_{ab} \vert,
\\
    v^{\mathrm{sig}}_{u} &\equiv \alpha_{u}
    \begin{cases}
        \sqrt{\frac{|P_{a} - P_{b}|}{\overline{\rho}_{ab}}}, & \qquad {\rm without\;gravity}
    \\
        |{\bf v}_{ab} \cdot \hat{{\bf r}}_{ab}|, & \qquad {\rm with \; gravity.}
    \end{cases}
\end{align}
Here $\alpha_{v,a}$ and $\beta_{v}$ are the usual linear and quadratic SPH viscosity parameters, respectively, while $\alpha_{\Delta \mathbf{v}}$ and $\alpha_u$ are both dimensionless coefficients of order unity. We use the convention that unbarred quantities with double indices indicate a difference term (e.g. $\mathbf{v}_{ab}^{\rm g} \equiv \mathbf{v}_a^{\rm g} - \mathbf{v}_b^{\rm g}$ and $u_{ab} \equiv u_a - u_b$), while barred quantities represent averages, e.g. $\overline{\rho}_{ab} \equiv \frac{1}{2} (\rho_a + \rho_b)$. Finally, $F_{ab}$ is defined such that $\nabla_a W_{ab} \equiv F_{ab} \mathbf{\hat{r}}_{ab}$.

\subsection{Positivity and boundedness of the dust fraction}
\label{sec:positivity}

During numerical evolution of the dust fraction, it is possible for the dust fraction to step outside of its physical bounds $0 \le \epsilon_{j} \le 1$. The implicit integrator from \citet{Elsender/Bate/2024} circumvents this issue, but has not been implemented in our version of \Phantom. Various schemes have been used in the past to address this issue \citep{Price/Laibe/2015,Hutchison/etal/2016,Ballabio/etal/2018,Hutchison/etal/2022}; however, in addition to being easy to implement, the multi-species adaptation of the parameterisation introduced by \citet{Ballabio/etal/2018} has been shown to be a good compromise between accuracy and computational speed
\begin{equation}
    S_{j} = \sqrt{\frac{\epsilon_{j}}{1-\epsilon_{j}}}
    \qquad {\rm or} \qquad
    \epsilon_{j} = \frac{S^2_{j}}{1 + S^2_{j}}.
    \label{eq:dustfrac_parameterisation}
\end{equation}
Note that individually constraining $0<\epsj<1$ does not provide the same upper bound on the weighted sum $\epsd$. This means that the parameterisation breaks down as $\epsd \to 1$. However, this is not a serious limitation since the one-fluid algorithm already fails in this regime due to the many terms with $1-\epsj$ in the denominator. It is also important to realise that $S_{j} \ne \sqrt{\rhodj/\rhog}$, which, by extension, means there is no easy way to extend the weighted-sum formalism to $S_{j}$.
Taking the time derivative of \cref{eq:dustfrac_parameterisation},
\begin{align}
    \frac{{\rm d}S_{j}}{{\rm d} t} =& \frac{(1 + S_{j}^2)^2}{2 S_{j}} \frac{{\rm d} \epsj}{{\rm d} t},
    \nonumber
\\
    =& - \frac{(1 + S_{j}^2)^2}{2 \rho S_{j}} \nabla \cdot\left[ \rho \frac{S_{j}^2}{1+S_{j}^2} (\deltavj - \epsd\deltav) \right],
\end{align}
we obtain a simple evolution equation for $S_{j}$ that directly utilises \cref{eq:final_dustfrac}. Although this form ensures that energy conservation is still maintained, the factor of $S_{j}$ in the denominator is problematic whenever $\epsj = 0$. Alternatively, we can remove the apparent singularity by factoring out an $S_{j}$ from within the divergence operator,
\begin{align}
    \frac{{\rm d}S_{j}}{{\rm d} t} =& - \frac{(1 + S_{j}^2)^2}{2}\left\{ \frac{1}{ \rho } \nabla \cdot\left[ \rho \frac{S_{j}}{1+S_{j}^2} (\deltavj - \epsd\deltav) \right] \nonumber \right.
\\
    & \left. \phantom{ - \frac{(1 + S_{j}^2)^2}{2}{\qquad}}  + (\deltavj - \epsd\deltav) \cdot \nabla S_{j} \right\},
\end{align}
which, after some manipulation, translates into the following SPH relation,
\begin{align}
    \frac{{\rm d}S_{ja}}{{\rm d} t} = & - \frac{(1 + S_{ja}^2)}{2} \sum_{b} m_{b} S_{jb} \left[\frac{(\deltavja-\epsda\deltava)}{\Omega_{a} \rho_{a}} \cdot \nabla W_{ab}(h_{a}) \phantom{\left( \frac{1+S_{ja}^2}{1+S_{jb}^2}\right)} \right.
    \nonumber
\\
    &
    \left. \qquad + \left( \frac{1+S_{ja}^2}{1+S_{jb}^2}\right) \frac{(\deltavjb-\epsdb\deltavb)}{\Omega_{b} \rho_{b}} \cdot \nabla W_{ab}(h_{b})\right] ,
    \label{eq:evol_dustfrac_param}
\end{align}
but the energy is no longer conserved to machine precision (errors in our test suite are on the level of $\sim 10^{-10}$) because the altered form of the derivative in \cref{eq:evol_dustfrac_param} no longer cancels exactly with the conjugate terms from the differential velocity and energy equations in \cref{eq:dedt}.

\subsection{Timestepping}
\label{sec:timestepping}

Strong drag regimes in multi-component flows typically impose prohibitive constraints on the hydrodynamic timestep in explicit solvers \citep{Laibe/Price/2012a}. Although the terminal velocity approximation made it easier to model small, tightly-coupled grains, timestep restrictions return whenever the dust decouples from the gas \citep{Price/Laibe/2015}. The only way to model both small and large grains in the same simulation is use an implicit drag algorithm. Using the implicit schemes from \citet{Laibe/Price/2014b} and \citet{Hutchison/etal/2016} as a base, we extend the formalism to account for multiple dust species. Efficient and scalable treatments of multifluid drag continue to be an active area of development, with recent GPU-oriented algorithms showing considerable promise \citep{Sewanou/Laibe/Commercon/2025}.

\subsubsection{Semi-analytic differential velocity relation}
\label{sec:timestep_deltav}

Much of the formalism is unchanged by the addition of more dust species. The analytic equation for the differential velocity is still obtained by assuming that the acceleration of $\deltavj$ in a barycentric fluid due to all non-drag forces is independent of the drag timescale $\tj$ and is approximately constant over each timestep, such that the system of equations is now represented in matrix notation by
\begin{equation}
    \frac{{\rm d} \deltavmat}{{\rm d} t} =  \matr{\Gamma} \deltavmat + \matr{A},
\end{equation}
where $\deltavmat$ and $\matr{A}$ are column vectors whose $j^{\rm th}$ elements are $\deltavj$ and $\mathbf{a}_{j}$, respectively. Here $\mathbf{a}_{j}$ is a placeholder for all terms on the right-hand side of \cref{eq:genmomentum_deltav} that do not involve $\tj$. Meanwhile, the elements of the drag matrix $\matr{\Gamma}_{jk}$ are given by
\begin{equation}
    \Gamma_{jk} = -
        \begin{cases}
            \frac{1}{\tj}\left( \frac{1}{\epsj} + \frac{1}{1-\epsd} \right),  & \qquad {\rm if \;} j = k
        \\
            \frac{1}{(1-\epsd) \tj}, & \qquad {\rm if \;} j \ne k .
        \end{cases}
    \label{eq:drag_matrix}
\end{equation}
Assuming the drag remains in the linear regime, such that $\tj$ is independent of $\deltavj$, the solution to this system of first order, ordinary differential equations with constant coefficients is
\begin{equation}
    \deltavmat(t) = {\rm e}^{\matr{\Gamma}t} \deltavmat(t_{0}) + (\mathbb{\matr{I}} - {\rm e}^{\matr{\Gamma}t}) \matr{\Gamma}^{-1} \matr{A},
    \label{eq:deltavsol_expmat}
\end{equation}
where we use $\mathbb{\matr{I}}$ to denote the identity matrix and $t_{0}$ is the time at the beginning of the timestep. 

Despite the simplicity with which we arrive at \cref{eq:deltavsol_expmat}, the hidden caveat is that our solution is written in terms of matrix exponentials. There are various methods for calculating the exponential of a matrix, but the process is particularly simple when the matrix in question can be diagonalised, in our case $\matr{\Gamma} = \matr{P} \matr{J} \matr{P}^{-1}$, with $\matr{J}$, the Jordan Form of $\matr{\Gamma}$, being a diagonal matrix. In this context, $\matr{P}$ is made up of the eigenvectors of $\matr{\Gamma}$ and the matrix exponential can be written as
\begin{equation}
    {\rm e}^{\matr{\Gamma} t} = \matr{P} {\rm e}^{\matr{J} t} \matr{P}^{-1}.
\end{equation}
Since the exponential of a diagonal matrix is just the exponential of the diagonal elements, the problem is reduced to solving for the eigenvalues and eigenvectors of $\matr{\Gamma}$.

All that remains now is to demonstrate the diagonisability of the drag matrix in \cref{eq:drag_matrix}. This is perhaps best shown by rewriting the drag matrix in terms of symmetric matrices, which are always diagonisable. To this end, we define the following matrices
\begin{align}
    & W_{jk} = -\frac{1}{(1-\epsd)\rho}
        \begin{cases}
            K_{j} \left(1+ \frac{1-\epsd}{\epsd_{j}} \right), &\qquad {\rm if \; } j = k
        \\
            \sqrt{K_{j} K_{k}}, & \qquad {\rm if \;} j \ne k
        \end{cases}
\\
    & \Psi_{jk} = 
        \begin{cases}
            K_{j}^{-1/2}, & \qquad {\rm if\;} j = k
        \\
            0, & \qquad {\rm if\;} j \ne k
        \end{cases}
\end{align}
such that we can decompose the drag matrix using the similarity transform $\matr{\Gamma} = \mathbf{\Psi} \matr{W} \mathbf{\Psi}^{-1} $. Then, inserting the diagonalisation of $ \matr{W} = \widetilde{\matr{P}} \matr{J} \widetilde{\matr{P}}^{-1} $, where the columns of $ \widetilde{\matr{P}} $ and the diagonal entries of $ \matr{J} $ are are the eigenvectors and eigenvalues of $ \matr{W} $, respectively, the eigenvector matrix for the original drag matrix is immediately apparent, i.e. $\matr{P} = \mathbf{\Psi}\widetilde{\matr{P}}$, while its eigenvalues are the same as for $\matr{W}$. Upon substitution of these results back into \cref{eq:deltavsol_expmat}, we obtain an analytic relation for $\deltavmat$ without matrix exponentials
\begin{equation}
    \deltavmat^{n+1} = \matr{P} {\rm e}^{\matr{J} \Delta t} \matr{P}^{-1} \deltavmat^{n} + (\mathbb{\matr{I}} - \matr{P} {\rm e}^{\matr{J} \Delta t} \matr{P}^{-1}) \matr{P} \matr{J}^{-1} \matr{P}^{-1} \matr{A}^{n}.
    \label{eq:implicit_deltav}
\end{equation}
Since this equation is only valid while $\matr{A} \approx {\rm constant}$ (i.e. over the timestep $n$), we have replaced $t \rightarrow \Delta t$ and added timestep labels as superscripts on the differential velocity and acceleration terms. 

\subsubsection{Semi-analytic energy relation}
\label{sec:timestep_u}

As we can see from the last term in \cref{eq:newusingle}, any change in the differential velocity affects the frictional heating of the gas due to drag. To properly account for the drag heating from all of the dust phases during a single timestep, we substitute the analytic solution for the differential velocity we just derived in \cref{eq:implicit_deltav} into the drag heating term and integrate from 0 to $\Delta t$
\begin{equation}
    \Delta u_{\rm drag} = \sum_{j} \frac{1}{(1-\epsd)\tj} \int_{0}^{\Delta t} \Delta V^2 \; {\rm d} t.
    \label{eq:implicit_udrag}
\end{equation}
The difficulty of completing this integral lies not in the functional time dependence, but from the coefficient matrices sandwiching the time dependent terms on either sides. Integration is performed term-by-term after first collapsing the matrix expressions to isolate the time dependence for each term. Fortunately, the structure of \cref{eq:implicit_deltav} is such that only three families of terms exist, based on the number of time-dependent exponential terms present (i.e. zero, one, or two). Once these general time-dependent terms have been integrated, e.g.
\begin{equation}
    I_{j} = \int_{0}^{\Delta t} \Delta V^2 \; {\rm d} t = I_{0,j} + I_{1,j} + I_{2,j},
    \label{eq:deltav2_integral}
\end{equation}
all that is left is to construct the constant coefficients with which they are paired
\begin{align}
    I_{0,j} &= t \sum_{i=1}^{\nu}  H_{j,i}^{2},
\\
    I_{1,j} &= 2 \sum_{i=1}^{\nu} H_{j,i} \sum_{\substack{k,l=1}}^{N}
         (F_{jkl,i} - G_{jkl,i}) \frac{{\rm e}^{\lambda_{k} \Delta t} - 1}{\lambda_{k}},
\\
    I_{2,j} &= \sum_{i=1}^{\nu} \!\sum_{\substack{k,l,\\m,q=1}}^{N} \!(F_{jkl,i} - G_{jkl,i})(F_{jmq,i} - G_{jmq,i}) \frac{{\rm e}^{(\lambda_{k}+\lambda_{m}) \Delta t} - 1}{\lambda_{k}+\lambda_{m}},
\end{align}
where $i$ is the coordinate index looping over $\nu$ spatial dimensions, $N$ is the number of dust phases embedded in the gas, $\lambda_{k}=J_{kk}$ are the eigenvalues of $\matr{W}$, and
\begin{align}
    F_{jkl,i} & = P_{jk} P^{-1}_{kl} \Delta V_{l,i},
\\
    G_{jkl,i} &= P_{jk} \lambda^{-1}_{k} P^{-1}_{kl} A_{l,i},
\\
    H_{j,i} &= \sum_{k,l=1}^{N} G_{jkl,i}.
\end{align}
Plugging $I_{j}$ into \cref{eq:implicit_udrag}, $u_{\rm drag}$ then replaces the first term in \cref{eq:final_energy} during the normal evolution of the internal energy.

\section{Benchmarking Tests}
\label{sec:benchmarking_tests}

We have generalised the dust tests in the \Phantom test suite \citep[see Section A.9 in][]{Price/etal/2018} to accommodate the full one-fluid equations, including adding a unit test to check that the derivative of the differential velocity is evaluated to within some tolerance of the expected value. After passing all of these sanity checks, we have further tested our implementation on three standard dust tests \citep[see, e.g.,][]{Laibe/Price/2011,Laibe/Price/2012a}: \textsc{dustybox}, \textsc{dustywave}, and \textsc{dustyshock}.

\subsection{\sc{dustybox}}
\label{sec:dustybox}

%
\begin{table}
	\centering
	\caption{Grain sizes ($\sj$) and dust fractions ($\epsj$) used in the \textsc{dustybox} test.}
	\label{tab:dustybox}
	\sisetup{table-format = 1.2,table-auto-round = true}
	\begin{tabular*}{0.75\columnwidth}
		{@{\extracolsep{\stretch{1}}}
			S[table-format=2.0]
			S[table-format=1.2e-1,scientific-notation=true]
			S[table-format=1.2e-2,scientific-notation=true]
		@{}} \toprule
		{$j$} 	&	{$\sj\,$[cm]}			&	{$\epsj$}				\\\midrule
		1	&	2.03476438e-04	&	9.87751904e-21	\\
		2	&	5.11109704e-04	&	2.48112060e-18	\\
		3	&	1.28384953e-03	&	6.23229298e-16	\\
		4	&	3.22488422e-03	&	1.56548044e-13	\\
		5	&	8.10054291e-03	&	3.93227809e-11	\\
		6	&	2.03476438e-02	&	9.87620274e-09	\\
		7	&	5.11109704e-02	&	2.47588610e-06	\\
		8	&	1.28384953e-01	&	6.02752025e-04	\\
		9	&	3.22488422e-01	&	9.37254100e-02	\\
		10	&	8.10054291e-01	&	4.05669352e-01	\\\bottomrule
	\end{tabular*}
\end{table}

The \textsc{dustybox} problem tracks the decay of the differential velocity between gas and dust in a system of uniform density, where the gas and dust are moving at different initial velocities.  It is an ideal test for our implicit drag solver since the only non-zero force is the drag between gas and dust. The analytic solution for this problem is derived in \cref{sec:timestep_deltav} and is used directly by the solver, so the results should be exact. 

Our setup consisted of a unit periodic box containing $8 \times 8 \times 8$ equally-spaced simulation particles at rest in the barycentric reference frame (i.e. $\vb = 0$), with equal amounts of gas and dust. All other gas and dust properties were set to values typically present in protoplanetary discs, the application where the dust module in \Phantom is most often used. For example, we set $\rhog = 5.33\times 10^{-11}\;{\rm g \, cm^{-3}}$ for the gas density and $\cs=1.49\;\rm {km\,s^{-1}}$ for the sound speed, values taken from a power-law disc model at a radius of $R=1\au$ \citep[see, e.g.][]{Hutchison/etal/2022}. Individual dust grains were assumed to be spherical with a constant intrinsic grain density of $3\;\rm g \, cm^{-3}$. Details of the drag coefficient calculation for physical grains in \Phantom can be found in Section 2.13 of \citet{Price/etal/2018}. We distributed the total dust mass across grains ranging in size from $\smin=1\mum$ to $\smax=1\cm$ according to the differential power-law distribution,
\begin{equation}
	\mathrm{d} \epsilon \propto s^{3-p} \mathrm{d} s, \qquad \mathrm{for} \quad \smin \leq s \leq \smax,
	\label{eq:MRN_distribution}
\end{equation}	
where $\mathrm{d} \epsilon$ is the differential dust fraction with respect to grain size $s$ and $p=3.5$ is the usual MRN power-law index for the grain size distribution \citep[e.g.][]{Mathis/Rumpl/Nordsieck/1977}. We discretised this distribution into 10 logarithmically spaced size bins, taking the mean value of each bin to be $s_j$, and integrated \cref{eq:MRN_distribution} across each bin to get the relative abundances for $\epsj$. Then, we normalised the relative abundances to be consistent with the desired total dust fraction (i.e. $\epsd = 0.5$). The discretised grain sizes and dust fractions used are summarised in \cref{tab:dustybox}.
Finally, we initialised all dust species with the same differential velocity of $\overline{\Delta v}_j=10^{-5}$ in code units ($29.78\;\rm cm \, s^{-1}$ in physical units) and the simulation is run until all dust species are equilibrated with the gas.

\begin{figure}
	\centering{\includegraphics[width=\columnwidth]{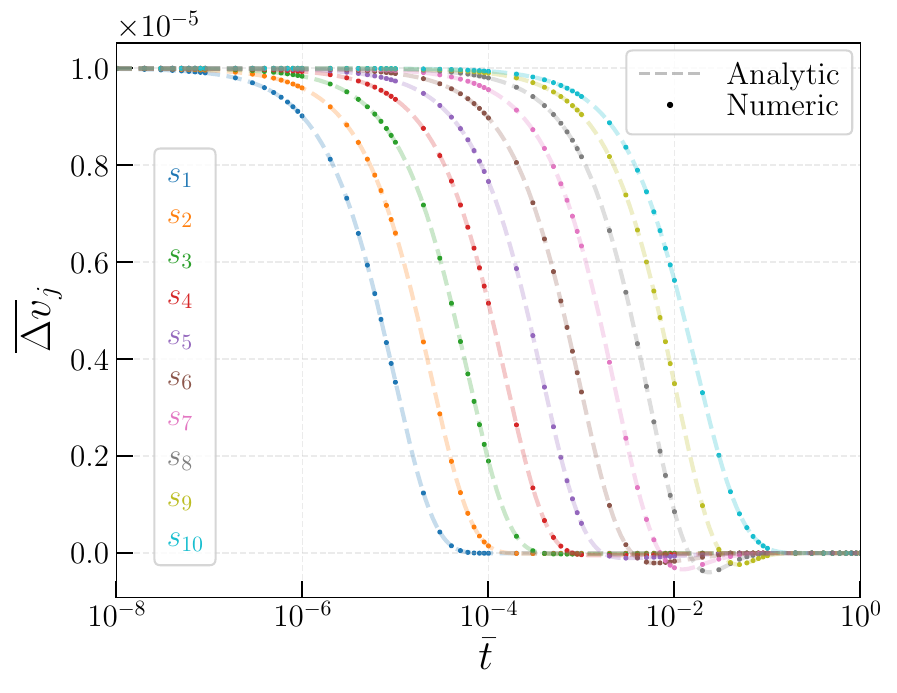}}
	\caption{Evolution of the differential velocity $\overline{\Delta v}_j$ as a function of time $\bar{t}$, both in code units, for the \textsc{dustybox} test assuming a dust-to-gas ratio of one. The dust is distributed across 10 species with grain sizes $s_j$ following a power-law distribution, as listed in \cref{tab:dustybox}. Colored points show the numerical results for each species, while dashed lines indicate the corresponding analytic solutions. Our implicit drag solver maintains accuracy across all drag regimes, even when using large time steps.}
	\label{fig:dustybox}
\end{figure}

\Cref{fig:dustybox} compares the numerical evolution of the differential velocity as a function of time (discrete points) against the analytic solution in \cref{eq:deltavsol_expmat} (dashed lines). All quantities are displayed in dimensionless code units, with length in units of $1 \au$, mass in solar masses and time units such that $\G = 1$. Tightly coupled grains equilibrate quickly with the gas, but since their dust fractions are so small, their effect on the gas evolution is negligible. It is not until the largest grain size with the highest dust fraction begins experiencing changes around $\bar{t} \sim 10^{-3}$ that the gas is significantly perturbed. While the small grains react immediately to these gas perturbations, larger grains that are still decelerating overshoot the gas and must be re-accelerated.

Since drag is the only active force in this test, there were no timestep restrictions. Here, we manually changed the time step after each time decade to demonstrate the implicit nature of the drag solver. However, we still recovered the correct solution even with much larger time steps -- including a single step covering the entire time domain. As expected, errors remained at the level of machine precision. We also tested configurations with different values of $\rho$, $\epsj$, $\epsd$, and $\deltavj$ and found similar results.

\subsection{\sc{dustywave}}
\label{sec:dustywave}

The \textsc{dustywave} test builds on the \textsc{dustybox} setup by introducing small sinusoidal perturbations to the otherwise uniform velocity, density, and energy fields. These perturbations generate linear sound waves in the gas, while the pressureless dust responds only through drag. This activates additional terms in the fluid equations while still neglecting small contributions from non-linear terms. The linearised equations describing linear multi-species dusty waves were derived by \citet{Laibe/Price/2014c} and are given later in \cref{eq:dustywave_lin_mass,eq:dustywave_lin_dustfrac,eq:dustywave_lin_momentum,eq:dustywave_lin_deltav}. 

For this test we placed $64 \times 12 \times 12$ simulation particles on a uniform, close-packed lattice in a periodic box with the long axis having a domain of $\bar{x} \in [-0.5,\,0.5]$ (dimensions in $y$ and $z$ were set to correspond to 12 particle spacings on the chosen lattice). The wave propagated along the $x$-axis with the initial conditions
\begin{align}
    \bar{v}_{\rm g} =&~ \bar{v}_{{\rm d}j} = A \sin{(2 \pi \bar{x})}, 
\\
    \bar{\rho}_{\rm g} =&~ 7\bar{\rho}_{{\rm d}j} =  1 + A \sin{(2 \pi \bar{x})},
\end{align}
where $A = 10^{-4}$. The density perturbation was initialised using stretch mapping \citep[see Appendix B in][]{Price/Monaghan/2004}. We performed this test using an adiabatic equation of state with $\bar{c}_{\rm s0} = 1$. We adopted a simple, constant $K$ drag prescription for the dust, choosing $\bar{K}_j =[10^3,\,10^2,\,10^1,\,10^0,\,10^{-1},\,10^{-2},\,10^{-3}]$ and the simulation was run until $\bar{t}=3$. All quantities are in code units. No dissipation or viscosity was applied to the system. 

\begin{figure}
    \centering{\includegraphics[width=\columnwidth]{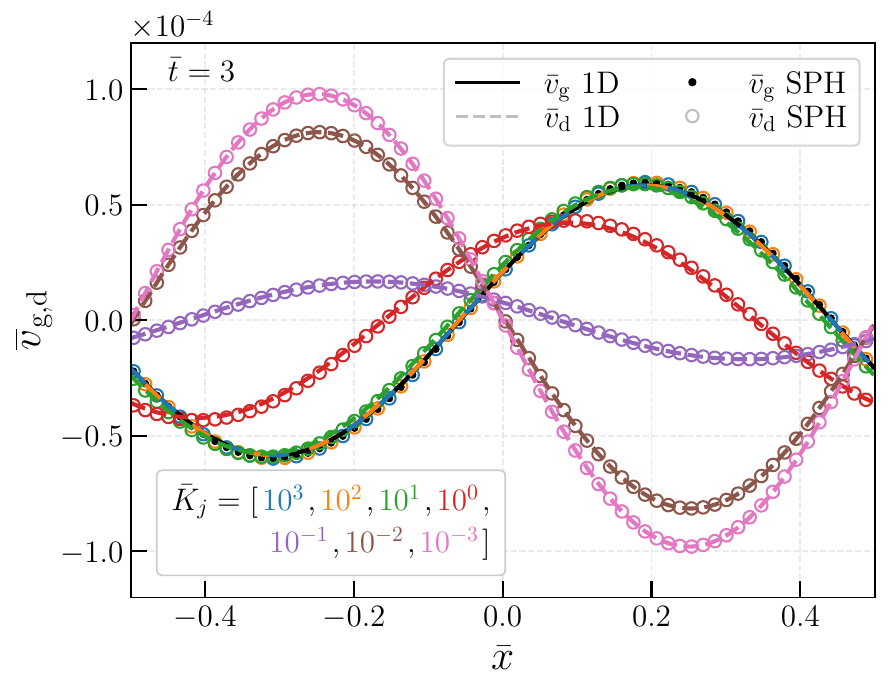}}
	\caption{Snapshot of the \textsc{dustywave} test at $\bar{t} = 3$ for seven dust species, each with a dust-to-gas ratio of $\,^{1}\!/\!_{7}$. Filled circles show SPH gas velocities, while open circles show dust velocities. Solid and dashed lines with colours matching the SPH markers give the corresponding gas and dust solutions from the high-resolution 1D spectral method. $L_2$ errors for the gas and dust components are $\lesssim2\%$ and converge toward the spectral solution with increasing space and time resolution.}
	\label{fig:dustywave}
\end{figure}
\Cref{fig:dustywave} compares the results from our SPH simulation (discrete points/circles) against a 1D spectral solution (solid/dashed lines) described in \cref{sec:dustywave_solution}. The 1D spectral solution was run with $1000$ grid points in $x$ using $50\,000$ uniform time steps so as to be considered `exact'. The $L_2$ errors are calculated according to
\begin{equation}
    \Vert e \Vert_{L_2} = \left[\frac{1}{N} \left( \frac{1}{f_{\rm max}^2} \sum_{i=1}^N \vert e_i
\vert^2 \right)\right]^{1/2},
\end{equation}
where $e_i = f_i - f_{\rm exact}$ is the error for each particle and $f$ is a placeholder for the quantity being tested. The normalised $L_2$ errors for the gas and dust are $\Vert e \Vert_{\rm g} = 1.81\%$ and $\Vert e \Vert_{{\rm d}j} = [1.81\%,\,1.81\%,\,1.83\%,\,1.99\%,\,0.86\%,\,0.08\%,\,0.08\%]$, respectively, consistent with a second-order integration scheme. These errors improve upon using higher resolution and/or forcing finer time steps. Again, we tested configurations with different values of $\rho$, $\epsj$, $\epsd$, and $\deltavj$ and found similar results.

\subsection{\sc{dustyshock}}
\label{sec:dustyshock}

%
\begin{figure}
    \centering{\includegraphics[width=\columnwidth]{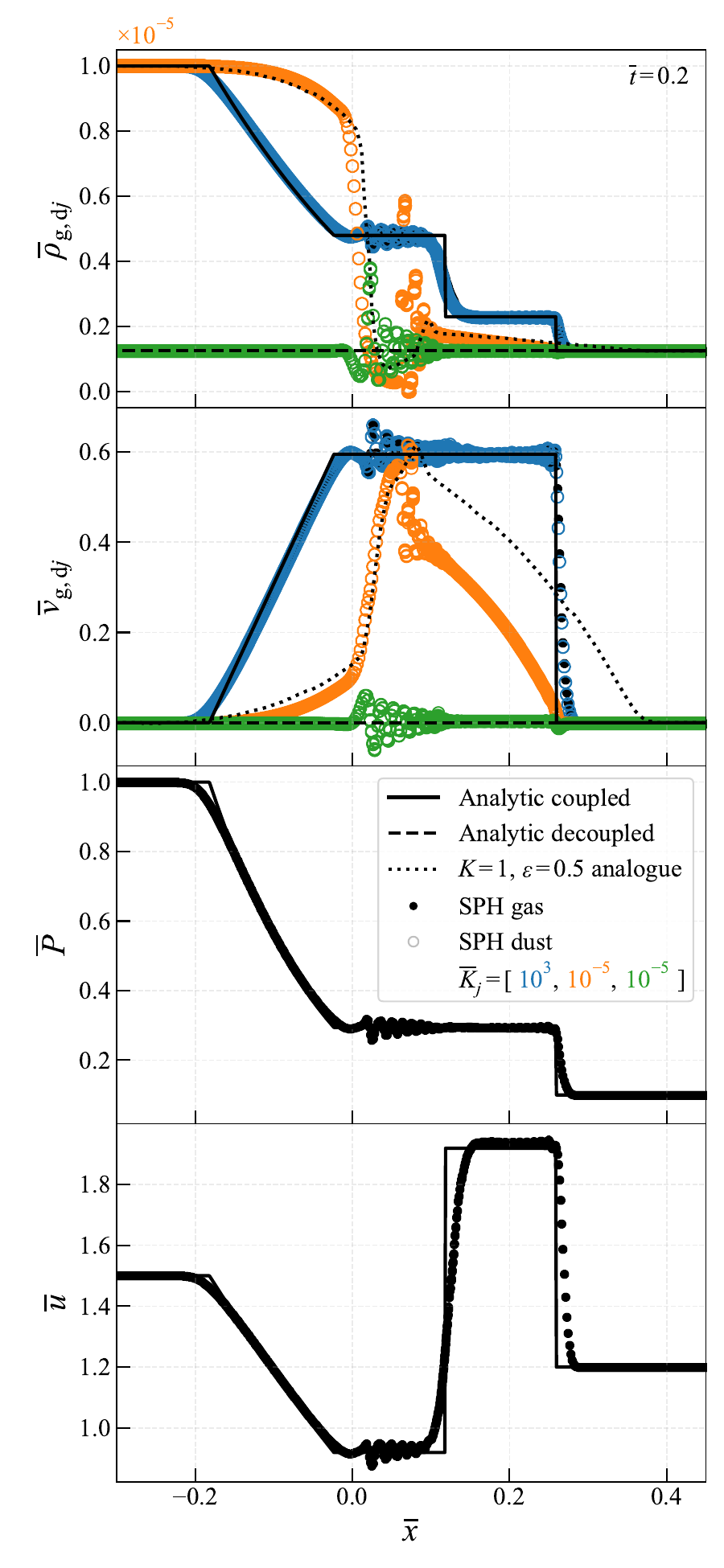}}
	\caption{Panels, from top to bottom, give the density, velocity, pressure, and internal energy in our \textsc{dustyshock} test at $\bar{t} = 0.2$ with three dust species.
    Filled circles represent SPH gas values, while open circles show dust values. The solid black line gives the analytic solution for the gas (filled black circles) and the $j = 1$ well-coupled dust (blue open circles), while the dashed black line is the analytic solution for the decoupled dust species (green open circles). Although we do not have an analytic solution for the weakly-coupled $j=2$ species, we compare it with the $K=1$ single-species two-fluid solution from \citet{Laibe/Price/2012a} with dust-to-gas ratio of unity. The first and third dust species closely follow their analytic solutions while the second dust species has the same expected morphology as its single-species analogue.}
	\label{fig:dustyshock}
\end{figure}

The \textsc{dustyshock} problem models a shock wave propagating through a gas-dust mixture, initiated by discontinuities in gas density, dust density, pressure, and internal energy. For this test, we employ the complete set of SPH fluid equations -- including viscosity, dissipation, and conduction terms. However, in weak drag regimes, we observed that extra dissipation in both the dust fraction and differential velocity was necessary to control post-shock oscillations that would otherwise grow unchecked. Our most effective configuration was achieved by adding the following non-conservative terms to \cref{eq:final_deltav,eq:evol_dustfrac_param}
\begin{align}
    & \frac{{\rm d}\deltavja^{\rm AD}}{{\rm d}t} =   \sum_b  m_b   \left[ \frac{q^{\rm AD}_{\Delta v,ja}}{\Omega_{a} \rho_{a}^{2}}  \nabla_{a} W_{ab}(h_{a}) 		+ \frac{q^{\rm AD}_{\Delta v,jb}}{\Omega_{b} \rho_{b}^{2}}  \nabla_{a} W_{ab}(h_{b})  \right] ,
    \label{eq:ad_deltav}
\\
    & \frac{{\rm d}S_{ja}^{\rm AD}}{{\rm d} t}  = 
      \sum_b  m_b   \left[ \frac{q^{\rm AD}_{S,ja}}{\Omega_{a} \rho_{a}^{2}}  F_{ab} (h_{a}) 		+ \frac{q^{\rm AD}_{S,jb}}{\Omega_{b} \rho_{b}^{2}} F_{ab} (h_{b})  \right] , 
    \label{eq:ad_dustfrac}
\end{align}
where we have defined
\begin{align}
    q^{\rm AD}_{\Delta v,ja} & \equiv \frac{1}{2} \rho_a v^{\rm sig}_{\Delta v, a} \left( \deltavja - \deltavjb \right) \cdot \mathbf{\hat{r}}_{ab},
    \label{eq:qad_dv}   
\\
    q^{\rm AD}_{S,ja} & \equiv \frac{1}{2} \rho_a v^{\rm sig}_{S, a} \left( S_{ja} - S_{jb} \right) \cdot \mathbf{\hat{r}}_{ab},
    \label{eq:qad_eps}
\end{align}
and the signal velocities $v^{\rm sig}_{S, a} =  v^{\mathrm{sig}}_{u,a}$ and $v^{\rm sig}_{\Delta v, a} = v^{\mathrm{sig}}_{v,a}$ have been recycled from the momentum and internal energy equations. While these terms help mitigate the oscillations, pronounced residual fluctuations remain, underscoring the need for additional refinement and investigation.

We initialised particles on a close-packed lattice with $256 \times 24 \times 24$ particles for $x \in [-0.5, 0]$ and $128 \times 12 \times 12$ for $x \in [0, 0.5]$. To maintain continuity across periodic boundaries, we set the number of particles in the $y$ and $z$ directions to the nearest multiple of 2 and 3, respectively, adjusting the $x$-spacing to preserve the correct density in each subdomain. Boundary conditions in the $x$-direction were enforced by tagging the first and last few particle rows as fixed boundary particles. We initialised the left and right states for the gas and dust to
\begin{align}
    (\bar{\rho}_{\rm g}, \bar{P}, \bar{u}) =& 
    \begin{cases}
        [1,\, 1,\, 1.5],  & {\rm if \;} x\le0
    \\
        [0.125,\, 0.125,\, 1.2], & {\rm if \;} x>0
    \end{cases}
\\
    \bar{\rho}_{{\rm d}j} =& 
    \begin{cases}
        [1,\, 10^{-5},\, 0.125], & {\rm if \;} x\le0
    \\
        [0.125,\, 1.25\times 10^{-6},\, 0.125], & {\rm if \;} x>0
    \end{cases}
\end{align}
with $\bar{v}_{\rm g} = \bar{v}_{{\rm d}j} = 0$. 
We used an adiabatic equation of state for the gas with $\gamma = 5/3$ and set constant drag coefficients for the dust: $\overline{K}_j = [10^3,\, 10^{-5},\, 10^{-5}]$. Although the second and third coefficients are identical, the effective coupling of each species is related to the drag time (\crefalt{eq:dragtime}), which is a function of both density and drag coefficient. Thus, when combined with their corresponding dust densities, these coefficients yield near-perfect coupling with the gas for the first species, intermediate coupling for the second, and almost complete decoupling for the third.

\Cref{fig:dustyshock} shows the gas and dust quantities at $\bar{t} = 0.2$. Since the $j=3$ species (green circles) is fully decoupled from the gas (solid black circles), its analytic solution is just its own initial conditions (dashed black lines). The density of the $j=2$ species (orange circles) is scaled up by $10^5$ for better comparison. While no analytic solution exists for the second species, its morphology closely resembles the $K = 1$ case from Figure 11 of \citet{Laibe/Price/2012a} (dotted black lines). Note, the prominent vertical offset in the post-shock velocity on the right is due to the slower shock propagation speed in our simulation, which is mass loaded with the well-coupled $j=1$ species (blue circles). Other, more subtle differences can be attributed to the negligible mass in our $j=2$ species that prevented any backreaction on the gas. However, the lack of backreaction in the second and third species was crucial for obtaining the combined analytic solution for the gas and $j=1$ species which assumes a gas-only solution with a modified sound speed  \citep[solid black lines;  see][]{Miura/Glass/1982}.

\subsection{\sc{dustysettle}}
\label{sec:dustysettle}

%
\begin{figure}
    \centering
    \includegraphics[width=\columnwidth]{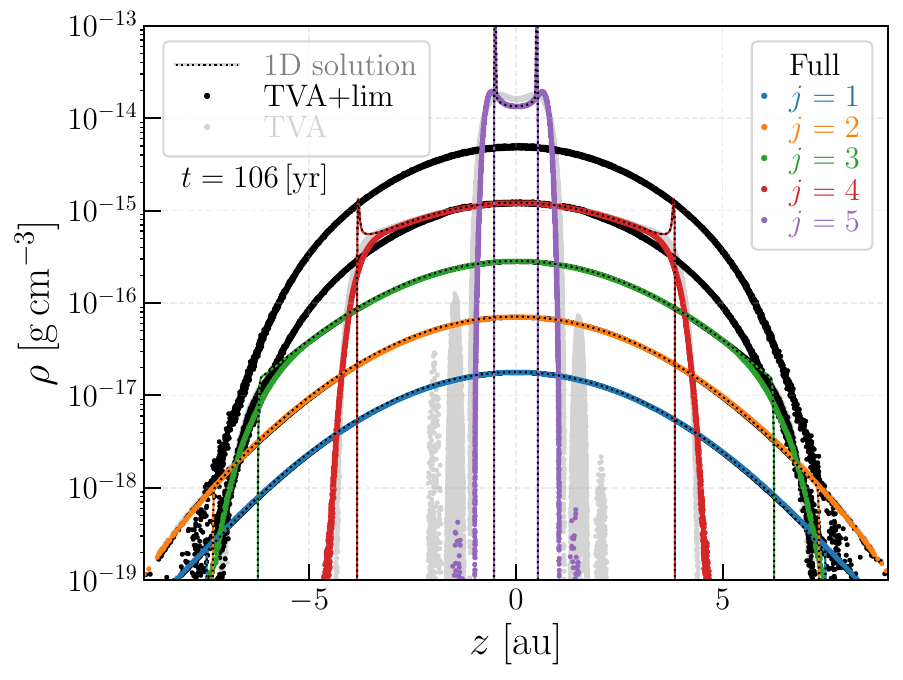}
    
    \caption{
    Results from the \textsc{dustysettle} test in a vertical column of gas and dust at $R=50,\au$ from a solar-mass star. {\it Left:} Coloured points show the full one-fluid simulation; black and grey points correspond to the terminal velocity approximation with a stopping-time limiter (TVA+lim) and without (TVA), respectively. Solid coloured curves with black dotted overlays show solutions from a high-resolution 1D solver. The SPH simulations do not resolve the sharp density peaks at the settling fronts because particle resolution is tied to total mass rather than dust mass. For the $s_{4}\approx 0.5\cm$ and $s_{5}\approx 8\cm$ grains, the stopping-time limiter in the TVA+lim model suppresses settling.    From this snapshot alone, it is tempting to conclude that the TVA simulation achieves similar accuracy to the full one-fluid solution even for large grains; however, comparing the time evolution of the two simulations reveals that this agreement is transient, occurring only as one solution overtakes the other (see \cref{fig:dustysettle_time_series}).}
    
    \label{fig:dustysettle_1d}
\end{figure}
\begin{figure*}
    \centering
    \includegraphics[width=\textwidth]{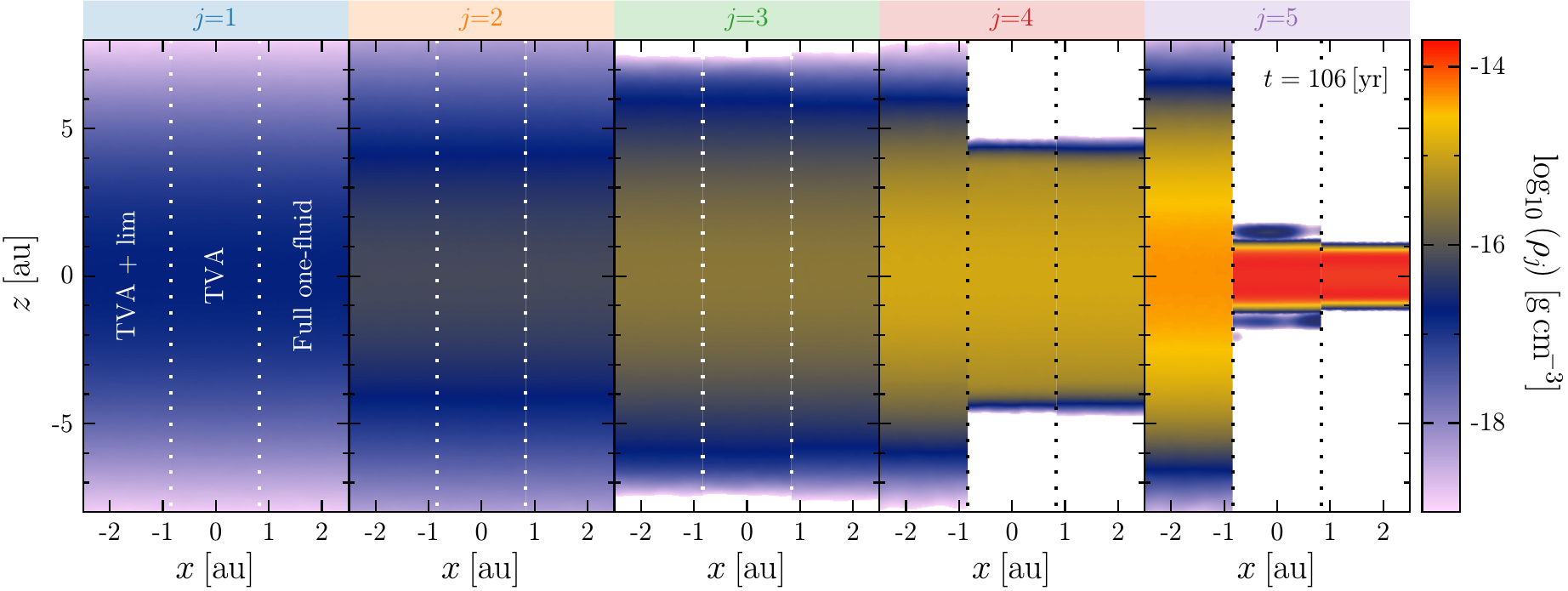}
    \caption{Rendered cross-sections of the five dust species in the \textsc{dustysettle} test. Each panel is subdivided into three columns, displaying from left to right the results from the TVA+lim, TVA, and full one-fluid simulations. All three methods show close agreement in the densities of the three smallest grain sizes. For the two largest grain sizes, however, the stopping-time limiter significantly reduces the settling speeds. The TVA alone appears more accurate, producing density distributions comparable to those of the full one-fluid simulation. As demonstrated in the right panel of \cref{fig:dustysettle_1d}, this agreement is largely coincidental since the TVA velocities are only accurate for a brief interval, around year 60, and differ from the full one-fluid solution both before and after this time.}
    \label{fig:dustysettle_2d}
\end{figure*}
\begin{figure*}
    \centering

    \begin{minipage}{0.5\textwidth}
        \centering
        \includegraphics[width=\textwidth]{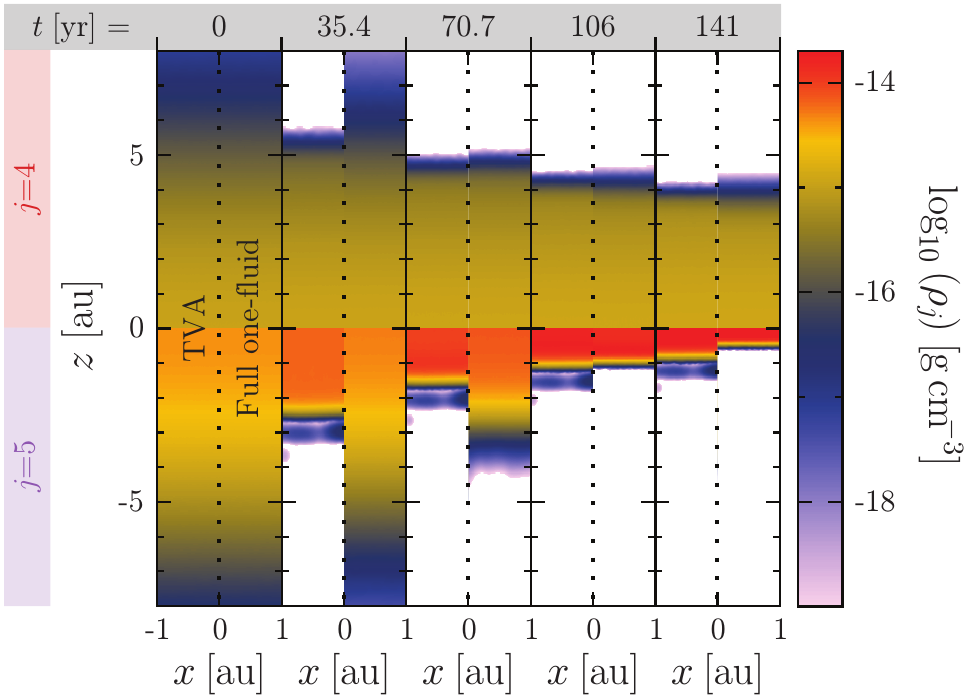}
    \end{minipage}
    \hfill
    \begin{minipage}{0.48\textwidth}
        \centering
        \includegraphics[width=\textwidth]{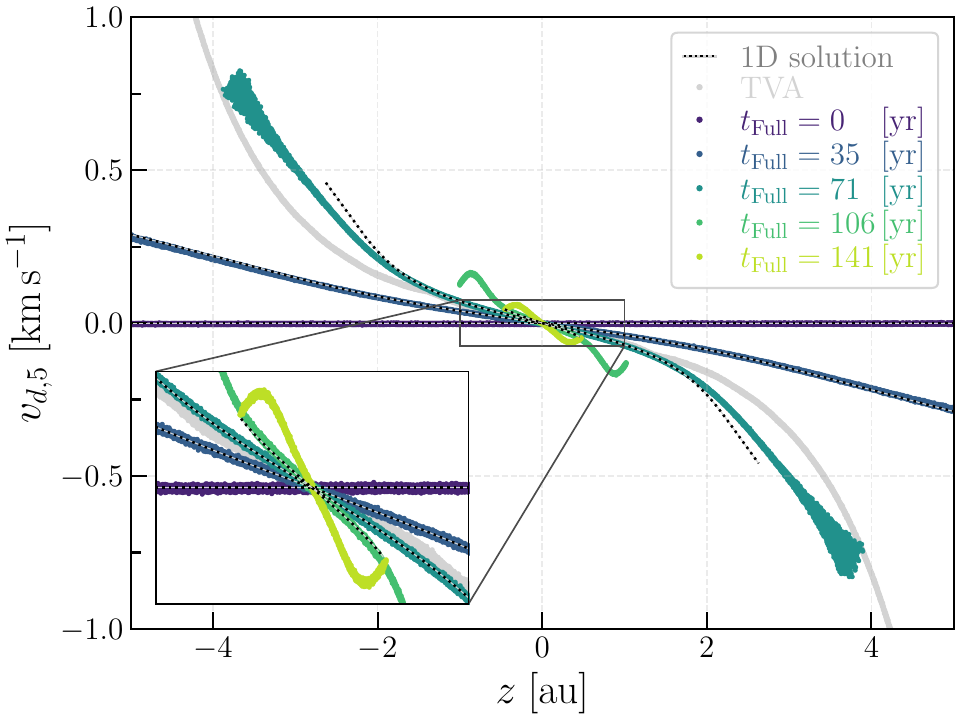}
    \end{minipage}

    \caption{
    {\it Left:} Rendered cross-sections of the two largest grains in the \textsc{dustysettle} test ($s_4$ for $z>0$ and $s_5$ for $z<0$), shown at $\sim35 \yr$ intervals. Each timestep is divided into two columns, comparing the TVA simulation (left) with the full one-fluid simulation (right). The $s_4$ grains in the upper atmosphere are poorly captured by the TVA for $|z| \gtrsim 5 \au$; however, the full one-fluid solution catches up by $t = 70 \yr$, after which both simulations evolve similarly, with only minor differences near the settling front. In contrast, the $s_5$ grains show significant deviations between the methods, except at $t = 106 \yr$, when the full one-fluid grains overtake and pass the TVA grains.  
    {\it Right:} Corresponding snapshots of the $s_{5}$ dust velocities for the dust distributions shown in the left panel. In the TVA simulation, dust velocities are fixed to the terminal velocity and remain nearly constant throughout the evolution. In contrast, dust in the full one-fluid simulation accelerates from rest and exceeds the local terminal velocity due to accumulated momentum during freefall from high altitudes. Residual gas oscillations from the relaxation phase are primarily responsible for the discrepancies between the full one-fluid and 1D solutions, the latter of which assumes the gas is stationary. Note the 1D Crank–Nicolson solver becomes unstable as the settling front sharpens, resulting in truncated profiles, and fails completely before the final snapshot.}
    \label{fig:dustysettle_time_series}
\end{figure*}

The previous tests have demonstrated the accuracy of the full one-fluid equations in different environments for a wide range of drag regimes. Now we would like to explore how the full one-fluid equations compare with the terminal velocity multigrain algorithm in \citet{Hutchison/Price/Laibe/2018} to better understand when the full equations are needed. Because the \citet{Hutchison/Price/Laibe/2018} implementation in \Phantom is almost always used in conjunction with the stopping-time limiter of \citet{Ballabio/etal/2018} for numerical stability in astrophysical environments \citep[see][for a way to avoid the limiter altogether]{Elsender/Bate/2024}, we run two sets of terminal velocity approximation simulations for the final two benchmarking tests: one with the limiter applied (hereafter TVA+lim) and one without (TVA). The \textsc{dustysettle} test is a good place to start because it isolates the vertical settling in a disc environment without the complication of radial drift, viscous expansion, or other disc-related phenomenon.

We setup our simulation using a thin, vertical (Cartesian) column of gas in near-hydrostatic equilibrium with an external acceleration in the form of
\begin{equation}
	\mathbf{a}_\mathrm{ext} = - \frac{\G M z}{\left(r^2+z^2\right)^{3/2}}\mathbf{\hat{z}},
	\label{eq:vertical_gravity}
\end{equation}
where $\G$ is Newton's gravitational constant, $M$ is the stellar mass, $r$ is the column's radial distance from the star, and $z$ is the vertical coordinate along the length of the column ($x$ and $y$ represent the two shorter dimensions of the column). The gas density of the column was given by
\begin{equation}
	\rhog(z) = \rhog{}_0 \, \mathrm{exp} \left[-\frac{z^2}{2H^2}\right],
	\label{eq:isothermal_density_profile}
\end{equation}
where we chose $H/r = 0.05$ assuming $r = 50\,\mathrm{au}$, giving a disc scale height of $H = 2.5\,\mathrm{au}$. We assumed an isothermal equation of state with $P = \cs^2 \rhog$, where $\cs \equiv H \Omega$ and $\Omega \equiv \sqrt{\G M / r^3}$, corresponding to an orbital time $t_\mathrm{orb} \equiv 2\pi/\Omega \approx 353\,\mathrm{yrs}$. We adopt code units with a distance unit of $10 \au$, mass in solar masses and time units such that
$\G = 1$. These choices give an orbital time of $\approx 70.2$ in code units.

Simulation particles were initially placed on a close-packed lattice using $32 \times 24 \times 111 = 85248$ particles in the domain $[x,y,z] \in [\pm H,\pm 0.6H,\pm 3H]$. Then, again using the method from \citet{Price/Monaghan/2004}, we stretched the particles in $z$ until we obtained the density profile given in \cref{eq:isothermal_density_profile}. We set the mid-plane density to $\rhog{}_0\approx 6 \times 10^{-13}\,\mathrm{g \, cm^{-3}}$ (or $10^{-3}$ in code units). We used periodic boundary conditions in all directions, but set the boundary in $z$ at $\pm 10 H$ in order to avoid periodicity in the vertical direction.

\begin{table}
	\centering
	\caption{Grain sizes ($\sj$) and dust fractions ($\epsj$) used in the \textsc{dustysettle} test}
	\label{tab:dustysettle}
	\sisetup{table-format = 1.2,table-auto-round = true}
	\begin{tabular*}{0.75\columnwidth}
		{@{\extracolsep{\stretch{1}}}
			S[table-format=2.0]
			S[table-format=1.2e-1,scientific-notation=true]
			S[table-format=1.2e-2,scientific-notation=true]
		@{}} \toprule
		{$j$} 	&	{$\sj\,$[cm]}			&	{$\epsj$}				\\\midrule
		1	&	1.25803588e-04  &	2.954510654748783E-05	\\
		2	&	1.99385250e-03	&	1.176211877132198E-04	\\
		3	&	3.16004325e-02	&	4.682583823765170E-04	\\
		4	&	5.00833104e-01	&	1.864170196958728E-03	\\
		5	&	7.93766977e+00	&	7.421395225413949E-03
        \\\bottomrule
	\end{tabular*}
\end{table}

We relaxed the gas density profile by running the code for 15 orbits followed by an additional $500 \yr$ with velocity damping. We then added five dust phases with a grain size distribution similar to \cref{sec:dustybox}, but with $\smax = 10 \,{\cm}$ and a total dust-to-gas ratio of 0.01 (specific grain sizes and dust fractions are given in \cref{tab:dustysettle}). \Cref{fig:dustysettle_1d,fig:dustysettle_2d} show the results after we continued the simulation for another three orbital time periods using each algorithm: TVA+lim (black points), TVA (grey points), and the full one-fluid equations (coloured points). Solid coloured lines with black dotted overlays give the solutions from the 1D solver described in Appendix A2 of \citet{Hutchison/Price/Laibe/2018}. Only small differences can be seen in the three SPH methods for the three smaller grain sizes, all of which are at high altitudes. On the other hand, the two largest grain sizes, $s_{4}\approx 0.5\cm$ and $s_{5}\approx 8\cm$ grains, exhibit a markedly different evolution in the TVA+lim simulation. A small portion of the discrepancy can be attributed to the known over-smoothing produced by the `direct second derivative' formulation in the terminal velocity approximation \citep[see Figures 8--9 and surrounding discussion in][]{Price/Laibe/2015}. However, the largest source for these differences is the stopping-time limiter, which forces the dust to behave as smaller particles with shorter stopping times, effectively slowing the settling (seen prominently in \cref{fig:dustysettle_2d}). Finally, note that the smoothing of the dust peaks at the settling front is a known consequence of the one-fluid method allocating resolution according to the total mass rather than the dust mass alone.

At first glance, the TVA simulation (i.e. without the limiter) appears to correct the inaccuracies seen in the TVA+lim densities for larger grain sizes. The full time evolution, however, reveals a more nuanced picture. The left-hand panel of \cref{fig:dustysettle_time_series} compares the density distributions of the two largest grain sizes in the TVA and full one-fluid simulations at $\sim 35 \yr$ intervals. In regions where the terminal velocity approximation breaks down -- particularly in the upper atmosphere -- the TVA solution quickly diverges from the full one-fluid result. Importantly, because the setup effectively restricts dust motion to the vertical direction, this early mismatch does not always lead to long-term differences in the density distribution.

This is most evident for the $s_4$ grains, which are sufficiently small to become overdamped in the exponentially increasing gas density. The transition to the terminal velocity regime for these grains occurs relatively high in the disc, with $\St\sim1$ at $z\sim6\au$ and $\St\sim0.1$ by $z\sim4\au$. The free-falling full one-fluid grains at high altitudes are thus able to catch up to the TVA settling front around $t\sim 70 \yr$, by which time they have become sufficiently coupled to the gas that the two simulations evolve in close agreement, with only minor differences in the shape of the settling front. The $s_5$ grains, by contrast, remain mismatched for most of the simulation, except for a brief period around $t \sim 106\yr$ when the full one-fluid grains overtake and pass the TVA grains. Owing to their roughly order-of-magnitude larger size, the $s_5$ grains settle closer to the critically damped limit since their Stokes numbers do not fall below unity until $z\sim2\au$. Their longer free-fall time at the start of the simulation both delays convergence with the TVA solution and allows the grains to accelerate beyond the local terminal velocity before drag becomes dominant.

This behaviour is illustrated in the right-hand panel of \cref{fig:dustysettle_time_series}, which shows the velocity evolution of the largest grain size at the same $\sim 35 \yr$ intervals. Importantly, dust in the terminal velocity approximation is instantaneously accelerated to the local terminal velocity where it remains for the duration of the simulation. Meanwhile, the dust in the full one-fluid simulation must accelerate from rest. Grains originating at high altitude can therefore acquire sufficient momentum during their descent to exceed the local terminal velocity at lower altitudes, producing a net faster settling rate than predicted by the TVA simulation. The full one-fluid equations capture this evolution well, as demonstrated by their agreement with the high-resolution 1D solver. Note, there are small discrepancies in the velocities -- most prominently at $t=71\yr$ -- which we attribute to residual gas oscillations in the SPH simulation. These oscillations explain why the full one-fluid velocities lag at $t=[35,\,71]\yr$ and lead at $t=106\yr$.

\begin{figure}
    \centering{\includegraphics[width=\columnwidth]{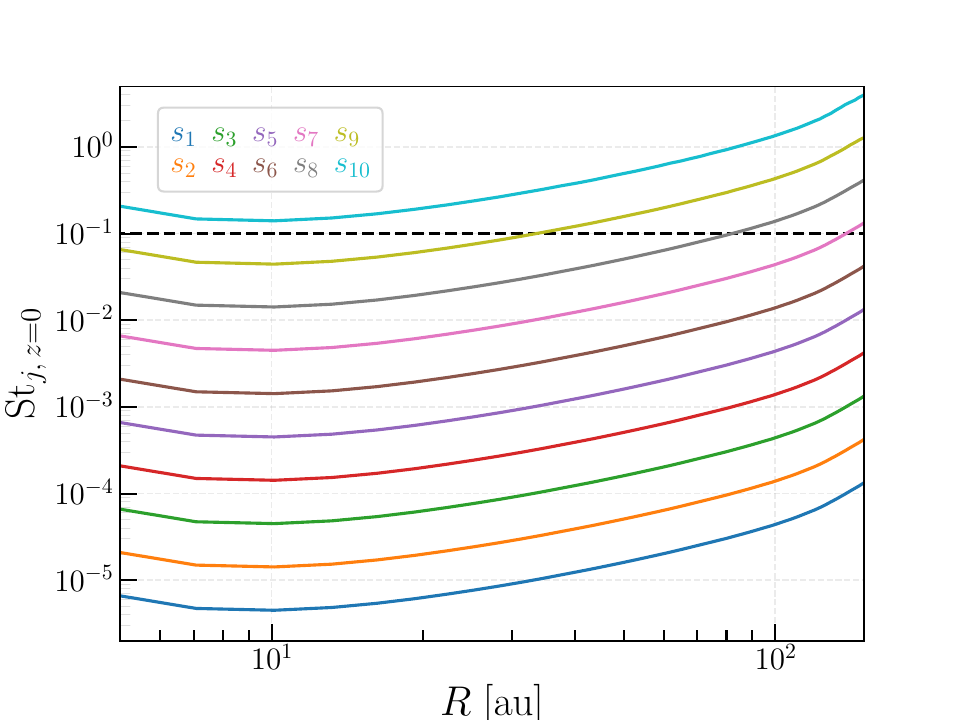}}
	\caption{Average initial Stokes numbers for dust species along the disc mid-plane in the \textsc{dustydisc} test. The horizontal black dashed line marks the threshold where errors in the terminal velocity approximation are expected to exceed $\sim10\%$. Since the Stokes numbers are inversely proportional to gas density, these profiles shift uniformly upward with increasing $\vert z \vert$.}
	\label{fig:dustydisc_St}
\end{figure}
\begin{figure*}
    \centering{\includegraphics[width=\textwidth]{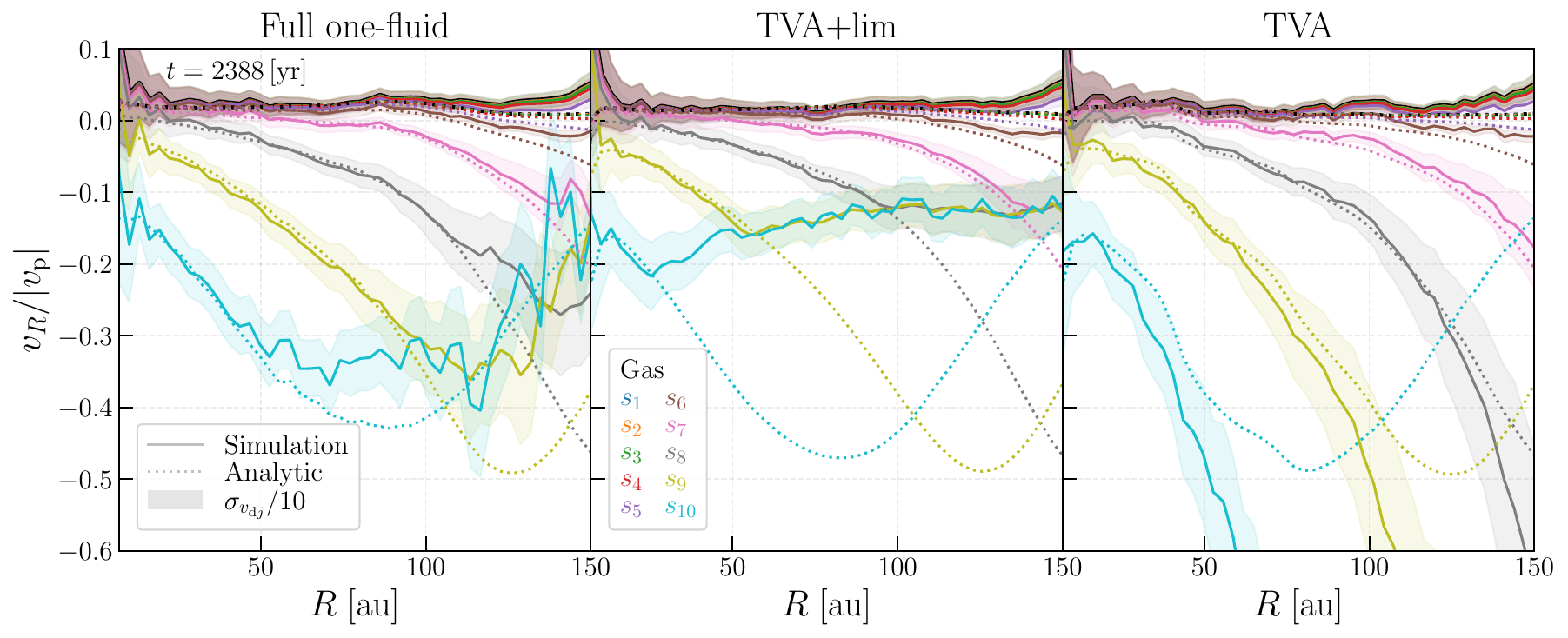}}
	\caption{Averaged radial gas and dust velocities along the disc mid-plane in the \textsc{dustydisc} test. From left to right the panels show the results from the full one-fluid, TVA+lim, and TVA simulations, respectively. Analytic solutions for each method are given as dotted lines. Velocities are normalised by the characteristic particle drift velocity ($v_{\rm p}$) from \cref{eq:particle_drift_velocity}, and shaded regions (scaled down by a factor of 10 for clarity) indicate standard deviations. Each method agrees well with theory for dust grains with Stokes numbers $\rm St\lesssim2$, beyond which both the TVA+lim and TVA diverge sharply from theory, albeit in very different manners. The TVA+lim velocities are capped by the stopping-time limiter and converge to a single velocity profile for large grains. On the other hand, the TVA velocities diverge to unphysically large magnitudes. These breakdowns highlight the terminal velocity approximation's fundamental inability to correctly model dust velocities for larger grains -- with or without the limiter.}
	\label{fig:dustydisc_vr_alone}
\end{figure*}
%

\subsection{\sc{dustydisc}}
\label{sec:dustydisc}

The previous test showed that the TVA+lim systematically underestimates settling velocities for weakly coupled grains, whereas the TVA tends to overestimate their velocities at early times and underestimate them at late times. However, in that setup, any drag-induced changes to the gas were immediately counteracted by pressure, minimizing the impact on other dust species. Despite the substantial differences in the settling rates of larger grains, the overall effect on the gas -- and, by extension, on smaller grains -- remained minimal. A natural next step is to compare these methods in a full disc simulation, where no true equilibrium state exists. Here, dust settles vertically and migrates radially, while the gas undergoes viscous accretion and expansion. Although pressure and gravity still enforce a quasi-equilibrium in the gas, the dust is free to follow a more diverse evolution. Importantly, even small differences can accumulate over time, potentially leading to divergent long-term evolution.

To test this theory, we performed three disc simulations with identical initial conditions using the same three methods as in the previous test, namely: TVA+lim, TVA, and the full one-fluid equations. We used `default' \Phantom setup parameters for the gas and a grain size distribution whose largest grain size moderately violates the terminal velocity approximation along the disc mid-plane (see \cref{fig:dustydisc_St}). More specifically, our setup consists of a vertically isothermal disc with radial power-law profiles for the gas surface density, sound speed, and disc aspect ratio
\begin{align}
    \Sigmag =&  \Sigmagref  \left(\frac{R}{\Rref}\right)^{-p} \exp{\left[-\left(\frac{R}{R_{\rm out}}\right)^{(2 - p)}\right]} \left(1 - \sqrt{\frac{\Rref}{R}}\right),
\\
    \cs =& \csref \left(\frac{\R}{\Rref}\right)^{-q},
\\
    h =& \frac{\Hg}{\R} 
    = h_0 \left(\frac{\R}{\Rref}\right)^{\frac{1}{2} - q}.
\end{align}  
Here we have used the subscript~$\reflab$ to denote reference values taken at a cylindrical distance $\Rref=1\au$ from the central star of mass $\Mstar = M_\odot$. The initial disc spanned $\R \in [1, 150] \au$, had a total disc mass of $\Mdisc = 0.05 \Mstar$, power-law indices of $p = 1$ and $q = 0.25$, and a reference aspect ratio $h_0 = 0.05$. The corresponding scaling parameters for the gas surface density and sound speed were $\Sigmagref \approx 912 \gram\cm^{-3}$ and $\csref \approx 1.5 \km \s^{-1}$, respectively. The gas viscosity was set via \Phantom's shock dissipation parameters $\alphaAV\approx 0.245$ and $\betaAV=2$ \citep[see][for more details]{Price/etal/2018}.

\begin{table}
	\centering
	\caption{Grain sizes ($\sj$) and dust fractions ($\epsj$) used in the \textsc{dustydisc} test.}
	\label{tab:dustydisc}
	\sisetup{table-format = 1.2,table-auto-round = true}
	\begin{tabular*}{0.75\columnwidth}
		{@{\extracolsep{\stretch{1}}}
			S[table-format=2.0]
			S[table-format=1.2e-1,scientific-notation=true]
			S[table-format=1.2e-2,scientific-notation=true]
		@{}} \toprule
		{$j$} 	&	{$\sj\,$[cm]}			&	{$\epsj$}				\\\midrule
		1	&   3.194201295916242e-05	&	2.444498094323023E-05	\\
		2	&   1.010095140015682e-04	&	4.347000629014016E-05	\\
		3	&   3.194201295916242e-04	&	7.730181714001872E-05	\\
		4	&   1.010095140015682e-03	&	1.374642297786892E-04	\\
		5	&   3.194201295916242e-03	&	2.444498094323021E-04	\\
		6	&   1.010095140015682e-02	&	4.347000629014017E-04	\\
		7	&   3.194201295916243e-02	&	7.730181714001870E-04	\\
		8	&   1.010095140015682e-01	&	1.374642297786892E-03	\\
		9	&   3.194201295916242e-01	&	2.444498094323023E-03	\\
		10	&   1.010095140015682e+00	&	4.347000629014016E-03	\\\bottomrule
	\end{tabular*}
\end{table}

We placed $500\,000$ gas particles in Monte-Carlo fashion and relaxed the disc for $3675 \yr$ to remove most of the perturbations from the initial particle placement. Normally, we would have initialised and relaxed the gas and dust together for this test, since mass loading the gas with dust introduces additional disturbances that must decay before we can measure radial velocities. Joint relaxation limits the time the disc can evolve away from its initial power-law profile, thereby providing more consistency with the analytic solution. However, to better isolate and interpret differences between the dust algorithms, we relaxed the gas disc first, reset the clock to $t=0$, and then added 10 dust phases to all particles within $R \le 150 \au$ following an MRN size distribution with a total dust-to-gas ratio of 0.01 (specific grain sizes and dust fractions are given in \cref{tab:dustydisc}). Note that large grains do not naturally populate the upper/outer regions of the disc, but this is often how SPH simulations without growth algorithms are initialised and thus provides an important test case.

Apart from the $s_{10}$ grains, our setup would be classified as a typical, low-resolution simulation that could be feasibly run under the terminal velocity approximation with `acceptable' errors. In practice, it is important to point out that due to numerical stability issues in the low-density regions of the disc (e.g. high altitude or radial outskirts), the terminal velocity approximation algorithm from \citet{Hutchison/Price/Laibe/2018} quickly crashes in disc environments, requiring the use of the stopping-time limiter from \citet{Ballabio/etal/2018} or the implicit integration scheme from \citet{Elsender/Bate/2024} for basic operation. Because we do not have the latter implemented, we took precautionary measures in the TVA simulation to prevent artificial dust formation by forcing the dust fractions of the three largest grain species to zero for $R \ge 165\au$. More stringent precautions would have been necessary if the disc had not first been relaxed or if the disc were not so smooth and quiescent.

To obtain the velocities for the \textsc{dustydisc} test, we binned the particles radially into 50 uniformly spaced bins and averaged the radial velocities both azimuthally and vertically within each bin. Because the radial velocities, densities and Stokes numbers change with $z$, we only included simulation particles within $|z| < \min(f_{\rm scale}H, z_{\rm cut})$ in our averaging process. We found that a fractional cut based on the disc scale height ($f_{\rm scale}H$) was more suitable for the inner disc, whereas a fixed height ($z_{\rm cut}$) was more effective in the outer disc, where $H$ is large. For grain sizes $s_{1\text{–}8}$, we adopted $f_{\rm scale} = 0.3$ and $z_{\rm cut} = 3\au$, while for the more efficiently settled $s_9$ and $s_{10}$ grains, we reduced these values to $f_{\rm scale} = [0.15, 0.15]$ and $z_{\rm cut} = [1.5, 0.3]\au$, respectively. Applying a uniform cut to all dust species led to over- or underestimation of some averaged radial velocities.

\Cref{fig:dustydisc_vr_alone} compares the calculated averaged radial velocities at $\bar{t} = 15\,000$ in code units ($\sim2388\yr$ in physical units) for the full one-fluid model (left panel), the TVA+lim (middle panel), and the TVA (right panel), together with their corresponding analytic solutions (dotted lines). All velocities are normalized by the characteristic particle drift velocity,
\begin{equation} 
    v_{\rm p} = -v_{\rm K} \left( \frac{H}{R}\right)^{2} \left| \frac{\partial \log{P_0} }{\partial \log{R}} \right|,
    \label{eq:particle_drift_velocity}
\end{equation}
where $P_0 = \rhog{}_0 \cs^2$ is the mid-plane pressure. The analytic solution for the radial gas and dust velocities comes from Equations 11 and 13, respectively, in \citet{Dipierro/etal/2018}. These expressions depend on $v_{\rm p}$, $\Stj$, $\rhog$, the radial velocity of the gas due to viscous torques, and the kinematic viscosity. Differences in the dotted curves between panels arise from variations in these locally computed quantities across the simulations. Using local values allows the analytic solution to adapt to each simulation; however, it does not fully compensate for deviations in the simulations from the underlying power-law assumptions of the model that develop during relaxation and evolution. For this reason, comparisons should be performed before the disc evolves too far from its initial state.

All three simulations show good agreement with the analytic solution for dust species $j=1$--$7$, as well as within limited regions of the disc for the three largest species. However, only the full one-fluid simulation reproduces the predicted rise and fall in migration speeds. Peak speeds occur at approximately the correct radial locations, but their magnitudes are systematically underestimated compared to the model. This discrepancy primarily arises from inconsistencies between the evolved disc and the idealised power-law profile. In particular, the measured velocities are sensitive to the mid-plane gas density used in the $v_{\rm p}$ scaling, whereas variations in $v_{\rm p}$ largely cancel in the analytic model. In fact, fitting a power-law profile to the gas density over the bulk of the disc (e.g. $R\in[20,\,100]\au$) for use in $v_{\rm p}$ leads to a slight overestimation of peak speeds, highlighting the sensitivity of the scaling to disc parameters. Additional deviations arise from residual disc oscillations, which predominantly affect the small grains, and from low-number statistics due to the stricter selection cuts applied to the two largest grains. Shaded regions indicate the standard deviations of dust velocities ($\sigma_{v_{{\rm d}j}}$, reduced by a factor of 10 for clarity). Thus, despite these deviations, the numerical and analytic solutions remain in qualitative agreement. Quantitatively, $L_2$ errors for the gas and five smallest grains are $\sim 40$--$50\%$, while the five larger grains have errors between $\sim 10$--$20\%$. These errors drop by almost half when restricting the analysis to $R \leq 100\au$.

The middle panel (TVA+lim) shows qualitatively similar trends but with significant departures from the analytic solution for $s_{8\text{–}10}$ at $R \sim [100,\, 50 ,\, 25]\au$, respectively. Beyond these breakpoints, the velocity profiles of these grains converge, reflecting how the stopping-time limiter caps the dynamics of large grains in order to prevent runaway accumulation of dust. This behaviour appears to be independent of the chosen vertical cut. The inability of the TVA+lim to capture the correct settling and migration velocities of the large grains explain the lack of disc processing compared to the full one-fluid.

\begin{figure}
    \centering{\includegraphics[width=\columnwidth]{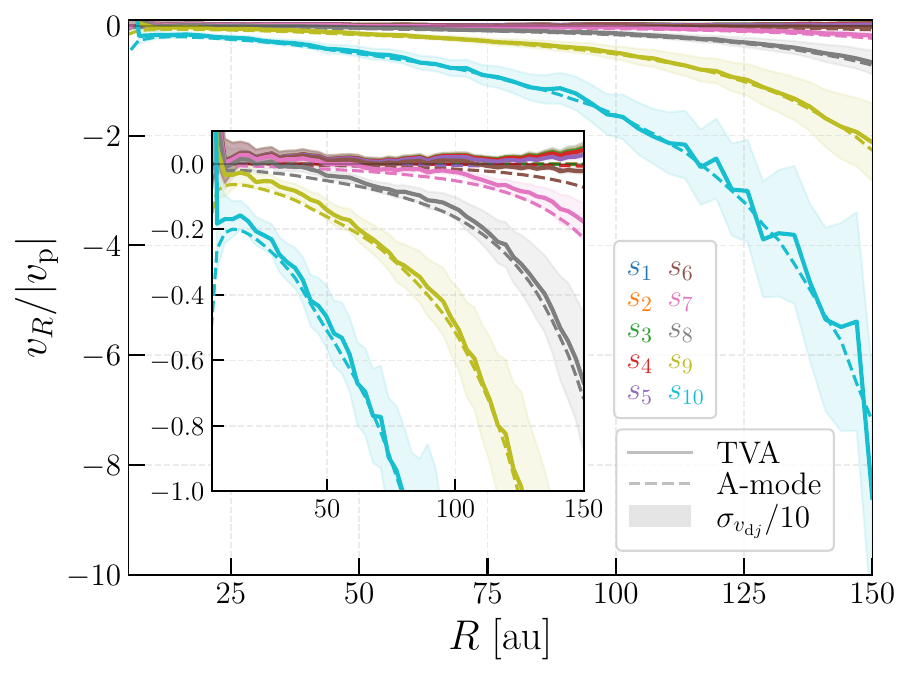}}
	\caption{The same TVA radial dust velocities shown in \cref{fig:dustydisc_vr_alone}, compared with the `A-mode' solution of \citet{Laibe/Gonzalez/Maddison/2012} for small, tightly coupled grains in an inviscid disc. The near-identical agreement shows that the TVA continues to follow the small-grain asymptotic solution even for large grains, leading to monotonically increasing drift speeds in the outer disc instead of the expected turnover that occurs for large Stokes numbers.}
	\label{fig:dustydisc_tva_vs_amode}
\end{figure}
\begin{figure*}
    \centering{\includegraphics[width=\textwidth]{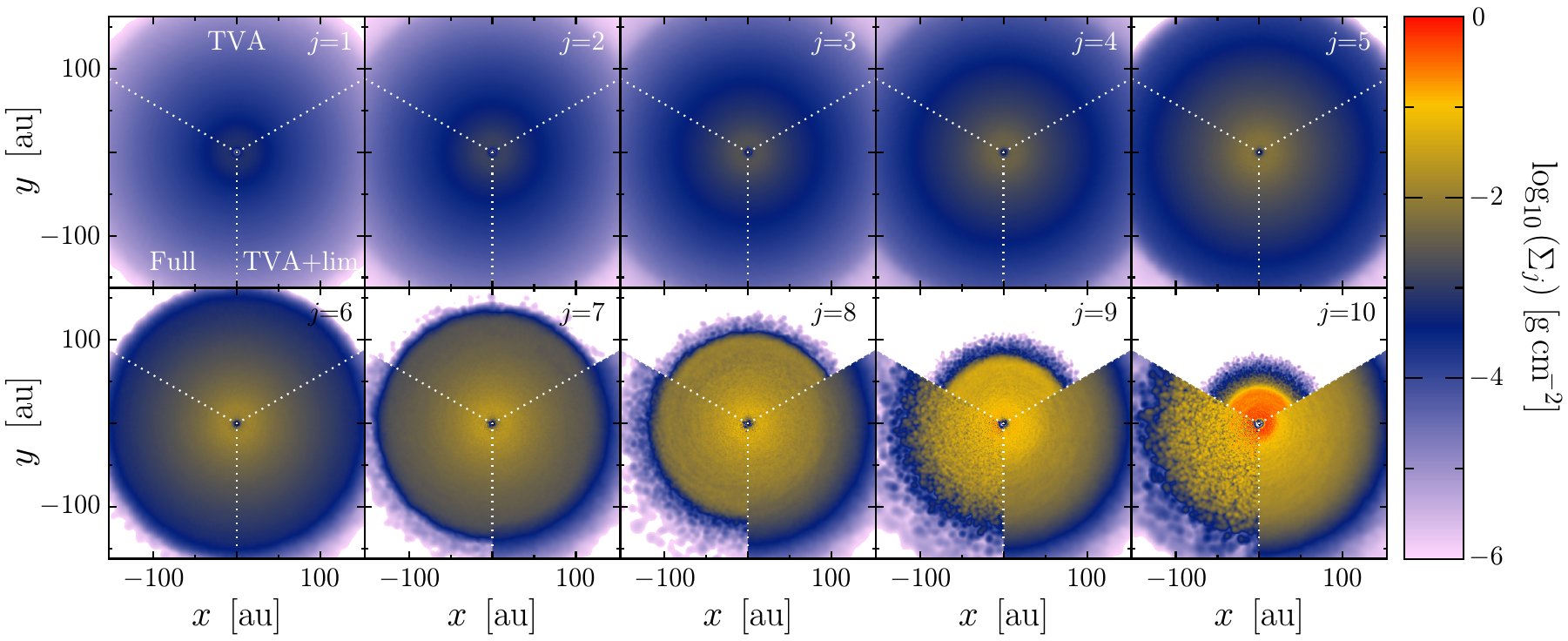}}
	\caption{Surface densities for all dust species. In each panel, the white dotted line separates the full one-fluid simulation (bottom left) from the TVA+lim and TVA simulations (bottom right and top, respectively). Significant deviations arise for species $j \ge 7$: the full one-fluid simulation exhibits dust distributions that are visibly more clumpy (reminiscent of two-fluid simulations) and more radially compact than in the TVA+lim case, though not as extreme as in the TVA without the limiter. The presence or absence of the stopping-time limiter therefore produces qualitatively different large-grain distributions, with implications even for moderately large grains in discs ($s \sim 0.1$--$1\mm$) that are commonly modelled using the terminal velocity approximation.}
	\label{fig:dustydisc_sigma_alone}
\end{figure*}
\begin{figure}
    \centering{\includegraphics[width=\columnwidth]{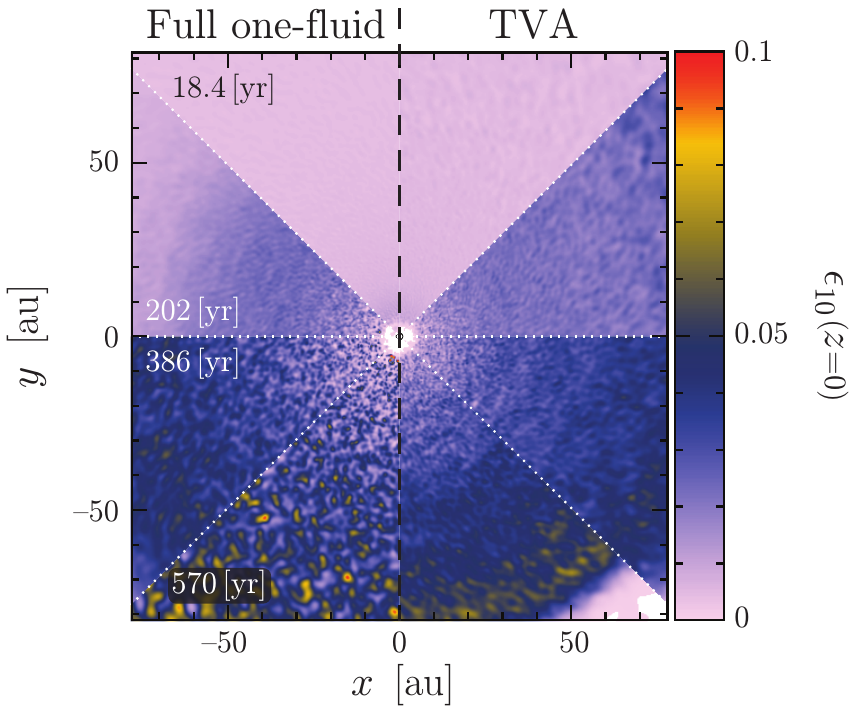}}
	\caption{Snapshots of $\epsd_{10}$ near the disc mid-plane at intervals of $184 \yr$, illustrating the onset of dust clumping. The left and right halves show the full one-fluid and TVA simulations, respectively. Although both simulations initially develop similar substructure, the TVA solution saturates after a few hundred years, while dust continues to concentrate in the full one-fluid simulation.}
	\label{fig:dustydisc_dustfrac_10}
\end{figure}

The right panel (TVA) also shows qualitative agreement with the analytic solution in the inner disc, but diverges markedly at radii $R \sim [140,\,100,\, 50]\au$ for grains $j=8$--$10$, respectively. In the absence of the limiter, the radial velocities follow a monotonically decreasing profile with large inward drift speeds in the outer disc. This behaviour is approximated well by the so-called `A-mode' of \citet{Laibe/Gonzalez/Maddison/2012}, derived for small, well-coupled grains embedded in an inviscid disc (see \cref{fig:dustydisc_tva_vs_amode}). In this regime, the radial velocity (their Equation 19) can be written succinctly as 
\begin{equation}
    v_{r,\textrm{TVA}} = - v_{\rm p}\St.
    \label{eq:vr_tva}
\end{equation}
When this scaling is extended beyond its regime of validity to larger grains (as we have done for this test), the linear dependence on Stokes number produces peak migration speeds that are more than an order of magnitude too large.

While the radial velocities are a useful diagnostic for verifying a method's accuracy, they represent only a single snapshot of the evolution. The cumulative impact of the dynamical differences between simulations is more clearly reflected in the dust distributions themselves. Although no exact solution is available for comparison, the density evolution provides a clear illustration of how differences between models can accumulate over time. To investigate this, we monitored the evolution of the gas and dust densities throughout the simulations to determine when the models began to diverge. Consistent with the results from the \textsc{dustysettle} test, the faster settling rates in the TVA and full one-fluid simulations began enhancing the mid-plane density of the $s_{10}$ grains within the first hundred years. Similar trends subsequently emerged in the smaller grains, but on progressively longer timescales. At the outer edge of the disc, the mid-plane density of the largest grains increased more rapidly in the TVA simulation than in the full one-fluid model, while the opposite trend was observed for the smaller $s_{8}$ and $s_{9}$ grains. Although the full one-fluid simulation generally exhibits slightly faster dust settling, as supported by the behaviour of the $s_{8}$ and $s_{9}$ grains, the early build-up of $s_{10}$ grains in the outer disc of the TVA run foreshadows the enhanced radial migration at large $R$ that becomes clearly evident later in the evolution. Meanwhile, the full one-fluid simulation diverged from the TVA+lim results in a different manner, with the largest changes occurring in the mid-plane densities of the inner ($\lesssim 10\au$) and outer ($\gtrsim 50\au$) disc. Differences in the surface density became more prominent around $t \sim 1000\yr$ and continued to grow with time.

\cref{fig:dustydisc_sigma_alone} shows the rendered surface densities for all dust species at $t = 2388\yr$, when the radial velocities were computed. Visible differences between the simulations are primarily confined to the four largest grain sizes. Relatively little evolution occurs in the TVA+lim case due to the enforced limits on settling and radial migration speeds. In contrast, the TVA simulation evolves rapidly, with the outer edge of the $s_{10}$ grain distribution contracting by a factor of $\sim 3$ in radius. The full one-fluid simulation generally lies between these two extremes, but with subtle differences in grain behaviour. For example, the outer edges of the $s_8$ and $s_9$ grain distributions undergo the largest contraction relative to the initial conditions. This is fully consistent with the predicted radial velocity profiles from \cref{fig:dustydisc_vr_alone}, as these grains are expected to attain the highest radial velocities for $R \gtrsim 100\au$. This provides additional evidence that the velocities in the full one-fluid simulation are being calculated correctly, particularly since the observed distributions reflect the cumulative effects of several thousand years of evolution rather than a single time-step.

\begin{figure}
    \centering{\includegraphics[width=\columnwidth]{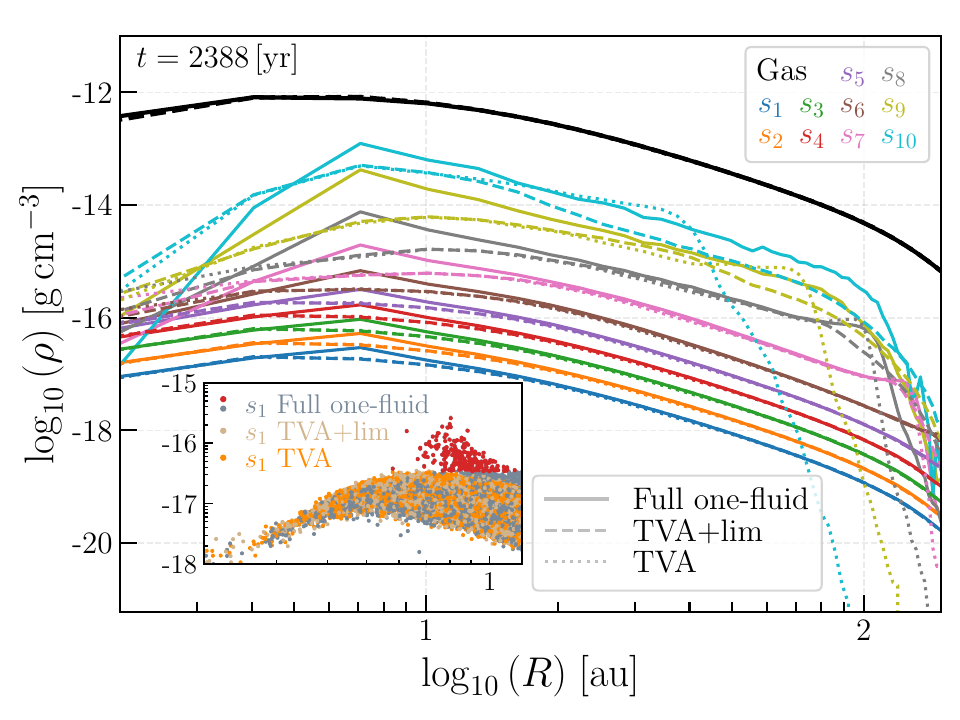}}
	\caption{Comparison of the averaged volume densities at $z=0$ for the gas (black lines) and dust (coloured lines) after $t = 2388\yr$ of simulation using the full one-fluid method (solid lines), the TVA+lim (dashed lines), and TVA (dotted lines). Differences between the three methods grow with grain size and are most pronounced in the inner and outer disc regions. In the outer disc, the truncated TVA profiles indicate elevated radial migration rates relative to the other methods, while enhanced dust accumulation in the inner disc ($R \sim 5\au$) of the full one-fluid simulation has already introduced deviations from the TVA+lim and TVA at smaller grain sizes. The latter of these effects is even more apparent in the small inset panel, which displays individual particle densities of the $s_{1}$ grains near the inner disc rim. Despite nearly identical accretion streams, the full one-fluid simulation exhibits elevated mid-plane dust densities and a pronounced peak (marked in red) at the dust trap location. A colour-coordinated plot of dust-to-gas ratios for the particles is provided in \cref{fig:dustydisc_dtg}.}
	\label{fig:dustydisc_rho}
\end{figure}
\begin{figure*}
    \centering

    \begin{minipage}{0.5\textwidth}
        \centering
        \includegraphics[width=\textwidth]{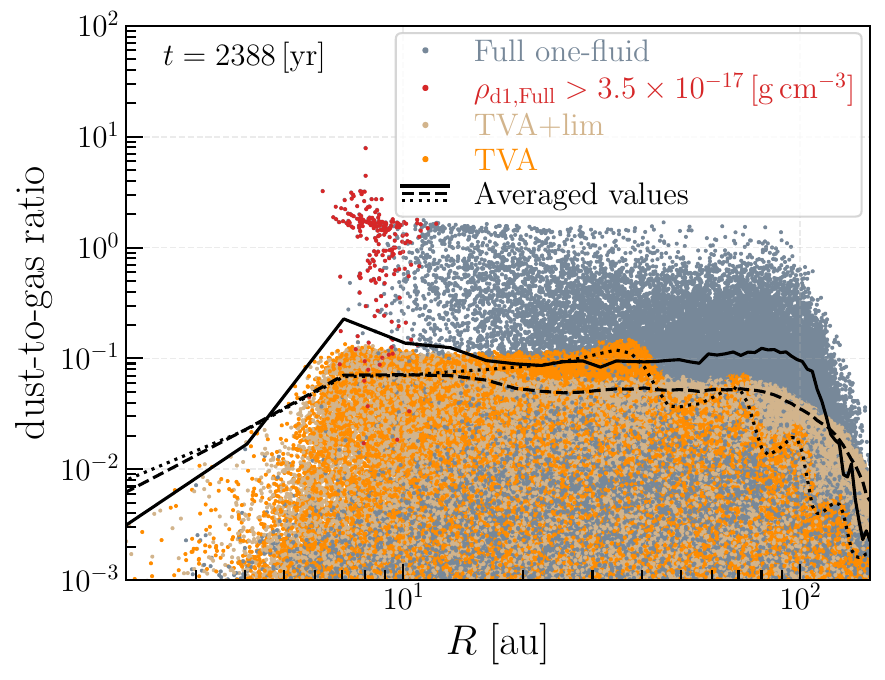}
    \end{minipage}
    \hfill
    \begin{minipage}{0.49\textwidth}
        \centering
        \includegraphics[width=\textwidth]{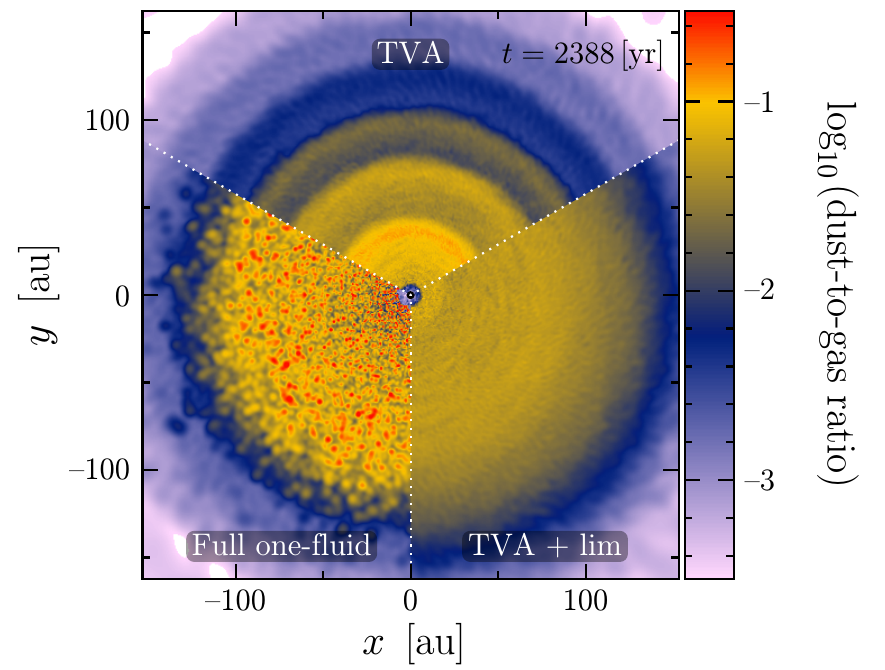}
    \end{minipage}
    
	\caption{ {\it Left:} Dust-to-gas ratios of particles in and around the disc mid-plane for the full one-fluid simulation (grey points), the TVA+lim (tan points), and TVA (orange points), with their respective averaged profiles shown as solid, dashed, and dotted lines. The broader spread of values in the full one-fluid simulation reflects its general settling and migration efficiency, but also its tendency towards dust clumping. Red points mark particles associated with the peak of the $s_{1}$ density distribution in the inset of \cref{fig:dustydisc_rho}, a structure only made possible by the strong backreaction from particles with dust-to-gas ratios exceeding unity. {\it Right:} Rendered cross-sections of the dust-to-gas ratio through each simulation’s disc mid-plane. The clumpiness responsible for the spread of values in the full one-fluid simulation in the left panel is clearly visible, as is the higher dust density. In contrast, the TVA+lim produces a much smoother dust distribution, but with an extended dust tail at the outer edge of the disc due to the dust limiter slowing the dust’s differential motion with the gas. Meanwhile, the TVA simulation shows four concentric dust rings at the peaks of the migration fronts of the four largest grain sizes, which are absent in the other simulations.}
	\label{fig:dustydisc_dtg}
\end{figure*}

A prominent feature of the full one-fluid results in \cref{fig:dustydisc_sigma_alone} is the development of a clumpy dust distribution for the $j=7$--$10$ grains. The TVA simulation also exhibits clumping near the steep density gradients trailing the migration fronts, but this behaviour is now well understood and can be controlled using the implicit solver of \citet{Elsender/Bate/2024}. More interestingly, closer inspection of the disc mid-plane at early times reveals that both methods initially develop remarkably similar inhomogeneities throughout the disc, with comparable growth rates and morphology (see \cref{fig:dustydisc_dustfrac_10}). This suggests that the onset of clumping is linked to the one-fluid formalism itself, likely due to its inability to represent multi-valued velocities for weakly coupled grains (although we have no clear proof of this). The two methods begin to diverge only after $\sim400 \yr$, when the TVA reaches a quasi-steady state while the overdensities in the full one-fluid simulation continue to grow. This behaviour reflects a fundamental difference between the two formulations: in the TVA, pressure-gradient-driven drift acts diffusively to smooth variations in the dust fraction, whereas the full one-fluid formulation permits purely advective evolution for weakly coupled grains and therefore lacks an equivalent smoothing mechanism. As a result, stochastic perturbations are retained more efficiently in the full one-fluid method, allowing overdensities to persist and amplify over time. Physically, we would expect differences in orbits within the unresolved dust distribution represented by each simulation particle to introduce an effective diffusion that counters unchecked growth. In practice, this suggests that simulations of large grains with the full one-fluid formalism will likely require some explicit regularisation of orbit crossing \citep[such as dust pressure or explicit diffusion; see, e.g.,][]{Tedeschi-Prades/etal/2025}. Since no such regularisation is included here, the long-term survival of the clumps is not considered a major concern, particularly because they do not appear to affect the measured radial migration rates.

\Cref{fig:dustydisc_rho} shows averaged mid-plane densities for the gas and dust at $t = 2388\yr$. The full one-fluid dust densities are systematically higher than those in the other simulations, particularly for the larger grains that experience efficient dust settling and radial migration. However, it is unexpected for the smaller grains to exhibit the same trend. It is likely that the enhancement of the smaller grains is (at least in part) a numerical effect arising from a clumping tendency of large grains in the full one-fluid simulation. Clumping of dust in the one-fluid formulation increases a particle's dust-to-gas ratio with a corresponding reduction in the gas mass fraction -- similar to the way sand displaces air in a jar. However, in astrophysical environments, the volumes are so vast that it would be virtually impossible to concentrate enough solids to displace sufficient quantities of gas to significantly affect the gas density. The only way to maintain a constant gas density with reduced gas mass is for the SPH particle to shrink, but this creates tension with the SPH requirement to maintain a suitable number of particle neighbours. In practice, the neighbour constraint tends to dominate, forcing displacement of the gas and thereby systematically increasing the relative dust concentrations of the other dust species. Moreover, post-process averaging of a clumpy distribution can also lead to small systematic offsets in mean values. Both effects are likely to have contributed here, especially in the inner disc, near the pressure bump at $\sim 5\au$.

In the outer disc, the four largest grains in the TVA simulation exhibit rapid radial migration, forming distinct migration fronts whose peak densities exceed those of the full one-fluid model. These fronts are analogous to the settling fronts in the \textsc{dustysettle} test and only form when the velocity increases sufficiently rapidly with distance for outer grains to overtake those at smaller radii (compare the TVA velocity profiles in \cref{fig:dustysettle_time_series,fig:dustydisc_tva_vs_amode}).
The general solution for the radial velocities (dotted lines in \cref{fig:dustydisc_vr_alone}) includes an additional factor of $(1+\St^2)^{-1}$ that is missing in \cref{eq:vr_tva}, such that the migration speed peaks and then declines with radius \citep[see Equation 13 in][]{Dipierro/etal/2018}. For the two largest grains, this peak lies within the disc, resulting in a decreasing velocity profile in the outer regions that does not support efficient front formation. For intermediate grains, however, the peak occurs outside the disc, yielding a monotonic increase in velocity across the disc interior, similar to the TVA case. Accordingly, the front location and magnitude for the $s_7$ and $s_8$ grains agree reasonably well between the two simulations. 
Finally, no fronts appear in the TVA+lim simulation because the stopping-time limiter caps the radial velocities, producing a nearly constant migration rate for large grains across much of the disc. As a result, the dust density profiles tend to closely follow that of the gas, which agrees well with the TVA simulation for all but the largest grains in the outer disc.

While the gas profiles exhibit only minor differences, backreaction from larger grains onto the gas is evident through its impact on smaller grains. For instance, the inset panel in \cref{fig:dustydisc_rho} shows the $s_1$ grain densities at the inner disc rim, where the density peak in the full one-fluid simulation (flagged with red points) has peeled away from the TVA+lim and TVA distributions. These simulation particles (not to be confused with the $s_1$ grains themselves) reside primarily in the mid-plane within the dust trap created by the gas pressure maximum. A large fraction of them exhibit elevated dust-to-gas ratios that exceed unity (\cref{fig:dustydisc_dtg}). The strong correlation between particles with the largest dust-to-gas ratios and those exhibiting enhanced densities in the smallest dust species likely indicates the presence of numerical artefacts. In particular, the measured enhancement is probably a combination of (i) numerical amplification associated with strong dust concentrations, as described above, and (ii) limited particle resolution in the inner disc, which isolates dust that would otherwise diffuse or stream into the accretion flow. Importantly, neither effect is intrinsic to the full one-fluid formulation itself. Thus, a TVA simulation encountering similar density and resolution conditions would be susceptible to the same numerical artefacts.

Not all of the observed behaviour should be attributed to numerical artefacts. The full one-fluid method still attempts to resolve the underlying dynamics, including genuine backreaction effects arising from the artificially enhanced dust densities. If the discrepancies were purely numerical, we would expect only systematic elevation in the dust fractions; instead, we also observe an inner dust cavity that is larger by $\sim1$--$2\au$. 
The dust in this region takes over as the dominant driver of evolution, accelerating the gas azimuthally and creating a shear layer with the dust-poor gas that continues to flow over the top/bottom. At least in our low-resolution simulation, this layer provides enough shielding from the surrounding fluid that even the smallest grains begin to accumulate with the rest of the dust migrating inwards along the mid-plane.

\begin{figure}
    \centering{\includegraphics[width=\columnwidth]{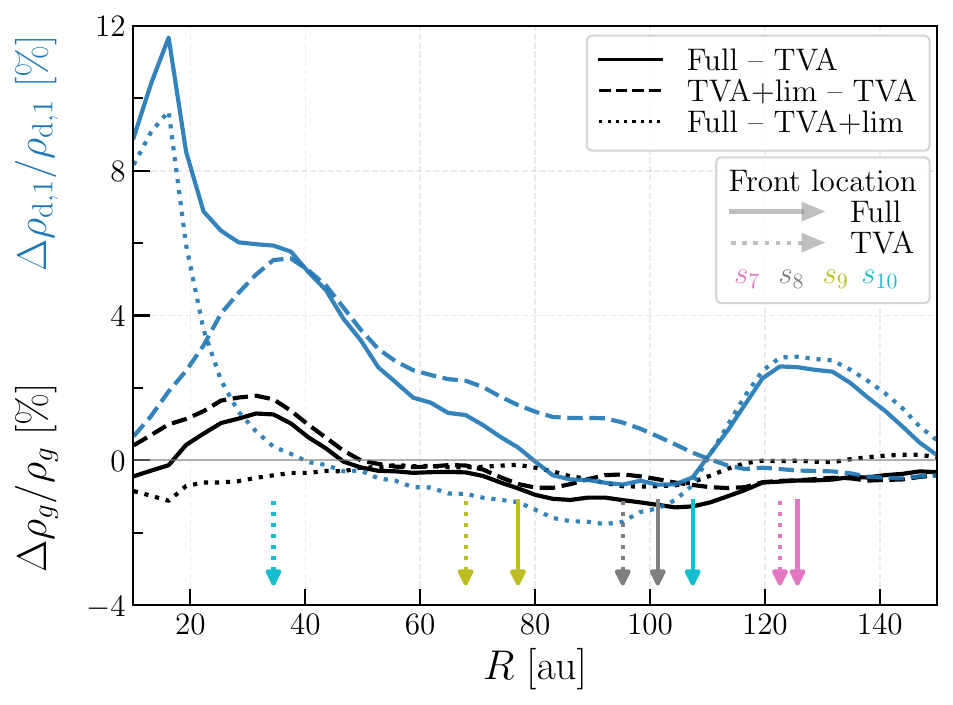}}
	\caption{Fractional differences in the gas and $s_1$ dust density profiles between the different methods at $t=2388 \yr$. The approximate migration-front locations of the four largest grains in the TVA and full one-fluid simulations are marked by vertical dashed and solid arrows, respectively. In the TVA simulation, the rapid migration of the $s_{10}$ grains induces strong gas backreaction, which in turn propagates to the other dust species. Differences between simulations are notably stronger in the dust than in the gas. Another prominent signal appears in the dust near the $s_7$ front locations in the outer disc. Overlapping perturbations from fronts within $R\in[70,\,110]\au$ obscures differences at intermediate radii.}
	\label{fig:dustydisc_gas_residuals}
\end{figure}

Determining how much of this behaviour is physical is difficult with the present setup. Nevertheless, these results demonstrate that inaccuracies affecting only one or two large dust species -- whether due to violations of the terminal velocity approximation or numerical artefacts -- can propagate through backreaction to all other dust species, including grains that should otherwise be well modelled by all methods considered here. This behaviour is also evident in other regions of the disc. For example, backreaction from the $s_{10}$ migration front in the TVA simulation produces $\sim 2\,\%$ fractional variations in the gas density and roughly three times that in the small dust species, with trailing effects extending over $\sim 60\au$ (see \cref{fig:dustydisc_gas_residuals}). Although these differences remain relatively small after only $2388\yr$, their cumulative effect over disc evolutionary timescales could ultimately lead to divergent evolution between simulations, affecting both the inferred dust distributions and any post-processed observables. If grain growth were included, these discrepancies could also propagate into the evolution of the grain size distribution as well.

%

\section{Discussion}
\label{sec:discussion}

Our results demonstrate that the full one-fluid approach reliably captures the complex interplay between gas and multiple dust species across a wide range of Stokes numbers. In our suite of benchmarking tests -- \textsc{dustybox}, \textsc{dustywave}, \textsc{dustyshock}, \textsc{dustysettle}, and \textsc{dustydisc} -- we find that the full one-fluid method reproduces the expected analytic solutions in every case. The inability of the terminal velocity approximation (with or without the limiter) to accurately model some of these regimes limits its functionality.

A major impetus for this study is the need for a framework capable of handling an evolving dust mass distribution driven by growth and fragmentation. Observational and numerical studies show that mm- to cm-sized grains are not only a natural by-product of coagulation in protoplanetary discs but also constitute a significant fraction of the solid disc mass. As we saw in the \textsc{dustydisc} test, failing to properly account for the dynamics of a large-grain population can alter both the evolution and the later interpretation of disc simulations. When using the terminal velocity approximation, we found that errors are not confined to the dust species that violate the approximation but can propagate to other species through their shared coupling to the gas, with the magnitude and location of these errors depending strongly on whether the stopping-time limiter is employed. The full one-fluid method showed similar behaviour, although in this case the errors arose from limitations of the one-fluid formalism itself rather than from a breakdown of the underlying equations.
What makes this problematic is that these errors remain inconspicuous until the results are compared against a more accurate method. 

One way to mitigate this issue is by supplementing the terminal velocity approximation with one or more two-fluid species \citep[see][]{Mentiplay/etal/2020}, thereby circumventing the problem entirely. 
However, incorporating growth and fragmentation into such a model would be formidable.\footnote{A notable exception might be when mass transfer is strictly one-way: for example, if coagulation and fragmentation act on the one-fluid simulation particles until the dust concentration exceeds a given threshold, at which point the dust is converted into two-fluid planetesimal particles.} In contrast, the one-fluid formalism naturally accommodates a changing mass distribution. Mass can be redistributed between dust bins without splitting or merging simulation particles, thereby maintaining consistent spatial resolution across all dust species -- even for those with very low abundances. Moreover, since all dust species share the same spatial coordinates, collision kernels have direct access to the full set of dust properties at a given location, eliminating the need for interpolation and enabling straightforward computation of differential velocities between species. Taken together, these features make the full one-fluid method the most promising framework for simultaneously resolving the dynamics of large grains and incorporating realistic models of dust growth and fragmentation.

There are two hurdles to evolving the mass distribution alongside the fluid equations in a 3D hydrodynamic simulation. First, the timescales for coagulation, fragmentation, and dust-gas dynamics can differ significantly, potentially disrupting momentum conservation in the hydrodynamic solver. Second, the representation of the mass distribution in different solvers can introduce subtle inconsistencies. Zeroth-order piecewise constant functions are likely the most compatible with hydrodynamic codes but are prone to numerical overdiffusion if the mass grid resolution is insufficient \citep[see][]{Grabowski/2022,Birnstiel/2024}. This issue is typically mitigated by using hundreds of dust bins, but such an approach would impose an excessive computational cost in 3D hydrodynamic simulations. To address this, \citet{Lombart/Laibe/2021,Lombart/Hutchison/Lee/2022,Lombart/etal/2024} have developed discontinuous Galerkin growth/fragmentation solvers, which use higher-order Legendre polynomial decompositions to reduce the number of dust bins while maintaining accuracy. However, this introduces a new challenge: the mass distribution must be repeatedly converted between the zeroth-order piecewise constant representation used by the hydrodynamic solver and the higher-order representation used by the discontinuous Galerkin solvers. Even if each solver individually conserves mass, their combination likely does not and thus requires special consideration.

Both hurdles present challenges for future development and will require carefully designed tests during the development phases. The full one-fluid method offers a more flexible framework for testing since the assumptions of the terminal velocity approximation prevent tests involving large initial differential velocities and weak drag regimes, such as the \textsc{dustybox} test. The \textsc{dustybox} test is particularly valuable in early development because it engages the hydrodynamic solver without moving the simulation particles. This controlled setup allows for a precise examination of relevant timescales, mass distribution conversion methods, mass and momentum conservation, and the optimal mass grid resolution and order number for the discontinuous Galerkin method to balance accuracy and computational cost. Moreover, among multispecies dust tests, the \textsc{dustybox} test is likely the only one simple enough to permit an analytic solution with a time-dependent mass distribution -- one that must be selected from the limited known analytic solutions for coagulation or fragmentation equations. Therefore, even if the terminal velocity approximation is ultimately preferred for production codes after incorporating grain size evolution, developing the correct integration procedure for the different solvers will likely be simpler and more efficacious with the full one-fluid equations.

While the full one-fluid method offers improvements over the terminal velocity approximation, it still has shortcomings. Chief among them is its inability to handle multivalued velocities, which occurs when dust grains follow intersecting trajectories \citep[clearly illustrated by the \textsc{dustyoscill} test in][]{Laibe/Price/2014b}. Although the terminal velocity approximation technically shares this weakness, it restricts itself to small Stokes number regimes where orbit crossings rarely occur, thereby avoiding the issue in practice (as long as the low-density regions can safely be ignored). In contrast, the two-fluid (i.e. dust-as-particles) method can simulate orbit-crossing dust grains without difficulty. This difference is analogous to the distinction between kinetic and fluid theories of gases. In the latter, the assumption of a single-valued velocity only makes sense when accompanied by a macroscopic pressure that encapsulates the microscopic velocity dispersion. While pressure is not typically associated with solids, dust grains nonetheless exhibit velocity dispersion -- whether due to collisions with gas molecules (e.g. Brownian motion), interactions with other grains (e.g. coagulation, fragmentation, bouncing), differential drag in a turbulent gas (e.g. during vertical settling or radial migration), or variations in orbital elements (e.g. eccentricity and inclination). A core problem with all existing one-fluid models is their treatment of dust as pressureless, despite the velocity spread these processes inherently produce. By not accounting for this spread in velocity, dust (especially large grains) tends to accumulate too efficiently, forming dense pockets (see \cref{fig:dustydisc_dtg}) and eventually producing unphysical dust fractions exceeding unity. One possible remedy is to include pressure-like diffusion terms, such as those motivated by turbulent transport \citep[e.g.][]{Klahr/Schreiber/2021,Binkert/2023,Tedeschi-Prades/etal/2025}, 
which can help limit over-concentration. However, even in the absence of turbulence, dust in particulate discs exhibits non-zero pressure arising from inclination and free eccentricity distributions \citep{Lynch/Lovell/Sefilian/2024}. These effects alone can produce significant deviations between the bulk motion of dust and the Keplerian orbits that individual grains follow. As there is still no consensus on how best to model dust pressure, and a full investigation lies beyond the scope of this paper, we leave this issue for future work.

The limitations discussed above do not imply that the terminal velocity approximation or the full one-fluid method are incapable of producing meaningful results. However, they do highlight the need for caution -- both in avoiding regimes known to be problematic for a given method and in not over-interpreting simulation outcomes. The \textsc{dustydisc} test in \cref{sec:dustydisc} serves as an instructive example: the rapid accumulation of dust near the mid-plane and inner disc of the full  one-fluid simulation, seen in \cref{fig:dustydisc_rho,fig:dustydisc_dtg}, and particularly the spike in small-grain density, is likely unphysical -- driven by the overly efficient concentration of large grains. Yet this behaviour is still numerically relevant. That the full one-fluid method was able to capture such subtle features in a relatively simple disc suggests there may be interesting physical effects yet to be uncovered, especially with broader grain-size distributions than are currently tractable under the terminal velocity approximation. Conversely, the seemingly more intuitive results produced by the TVA+lim and TVA simulations in the same region are not necessarily indicative of greater accuracy, as illustrated in the middle and right panels of \cref{fig:dustydisc_vr_alone}. Producing plausible outcomes from flawed calculations can be just as misleading. Much of the uncertainty discussed here arises from using an unphysical grain-size distribution (and in the case of the terminal velocity approximation, also conflicts with the method’s underlying assumptions). In reality, grain growth and fragmentation would vary with local disc properties and timescales, with the largest grains first appearing near the mid-plane. Incorporating a self-consistent grain-size distribution shaped by growth and fragmentation would not only improve the realism of such simulations but may also help prevent the full one-fluid method from reaching unphysical dust concentrations.

Two final numerical challenges are worth discussing. First, in the \textsc{dustyshock} test, we were unable to find an artificial dissipation switch that fully suppressed the post-shock oscillations in the dust. Although strong oscillations developed only when a weakly coupled dust phase was present, they quickly propagated to the other phases. Even when reverting to a single dust species and following the prescription of \citet{Laibe/Price/2014b}, our post-shock oscillations remained noticeably larger than those reported in their work. Ultimately, we obtained our best results by (i) applying dissipation to both the differential velocities and dust fractions, and (ii) following the recommendation of \citet{Laibe/Price/2014b} to omit artificial viscosity terms from the $\fg$ term in them$\deltavj$ evolution equation. However, applying this same prescription in the \textsc{dustydisc} test severely affected the dynamics, leading to unphysical behaviour -- including high-speed ejection of particles from the disc, likely caused by non-conservation of momentum or energy. For this reason, we ran the \textsc{dustydisc} test without either of the above dissipation methods used in the \textsc{dustyshock} setup. It is possible that better dissipation switches for the dust could be constructed using the total time derivative of the divergence of individual differential velocities and/or dust fractions, similar to the \citet{Cullen/Dehnen/2010} shock indicator we use for the gas. In any case, it is clear from these tests that our approach to shock capturing in the dust is flawed and warrants further investigation.

The second numerical challenge is computational cost. In addition to the standard evolution equations used in the terminal velocity approximation, the full one-fluid method must additionally evolve the three components of the differential velocity for each dust phase, increasing both memory usage and computational load. The greatest cost, however, comes from the implicit drag solver required to integrate the drag terms in the differential velocity and energy equations. While still faster than an explicit scheme, this solver introduces substantial overhead due to the eigenvalue decomposition of the drag matrix and the construction of integrated energy terms for every simulation particle. In the \textsc{dustydisc} test, intervals between snapshots were typically five to ten times longer than those for the terminal velocity approximation. When accounting for future additions such as grain growth and fragmentation -- likely to require several tens of dust bins -- the overall computational burden will grow substantially. Codes designed for GPU acceleration, such as the new Shamrock code \citep{David-Cleris/Laibe/Lapeyre/2025}, may offer a more suitable framework for such demanding simulations.

\section{Conclusions}
\label{sec:conclusions}

In this work we have presented the first SPH implementation of the full one-fluid, multi-species dust-gas equations derived by \citet{Laibe/Price/2014c}, correcting key errors in the continuum derivation along the way. Our method is comparable to the multi-species terminal velocity approximation introduced by \citet{Hutchison/Price/Laibe/2018} but without the constraints on the Stokes number, thereby allowing us to simulate both tightly and loosely coupled grains in a wider range of initial conditions and scenarios. Our implementation makes it possible to avoid prohibitive timestep restrictions for the tightly coupled grains using our implicit drag solver that maintains the mass, energy, and momentum conservation inherent to the SPH formalism. This combination of accuracy, efficiency, and flexibility in a single framework represents a significant improvement upon previous methods and opens up new avenues for exploring the physics of dusty astrophysical systems.

Our benchmarking tests demonstrate that the full one-fluid model accurately reproduces analytic solutions across all drag regimes, including capturing the effects of backreaction from each dust species onto the gas and the resulting indirect coupling this creates between dust species. In the final two tests (\textsc{dustysettle} and \textsc{dustydisc}), we directly compared our results with the terminal velocity approximation with (TVA+lim) and without (TVA) the stopping-time limiter from \citet{Ballabio/etal/2018} to highlight where its assumptions break down and how resulting errors propagate over time. Crucially, we find that such errors are not confined to the dust species for which the assumptions fail but are transmitted to other species via their coupling to the gas. This statement holds whether or not the limiter is applied, but the details of how the errors propagate differ between the two formulations. While this error communication is currently mediated only through backreaction, future models that couple coagulation and fragmentation with hydrodynamics will link dust species directly, further amplifying such effects. Given that mm- and cm-sized grains are natural by-products of coagulation and tend to dominate the local dust mass once formed, the full one-fluid method will be essential for achieving accurate results in these more realistic simulations. But even beyond accuracy, the extended flexibility of the full one-fluid method also makes it a powerful framework for development and testing. It is inherently well suited to growth and fragmentation models, as it (i) allows mass to be redistributed between dust bins without modifying the underlying particle distribution, and (ii) keeps all species spatially collocated, ensuring consistent resolution and efficient evaluation of inter-species dynamics. These features position the full one-fluid method as a practical and physically motivated framework with which to build next-generation models of dust evolution in astrophysical environments.

While this work represents a significant step forward in modelling gas–dust mixtures with SPH, several limitations remain. Foremost among these is the one-fluid method’s inability to handle multivalued dust velocities arising from orbit crossing or turbulent dispersion -- an issue rooted in the inconsistency between the single-valued nature of the fluid approximation and the assumption of a pressureless dust phase. On large scales, this prevents the one-fluid method from being used to model interpenetrating dust streams, such as dust oscillating about the disc mid-plane. On smaller scales, it manifests as local dust clumping, the early development of which we observe in both the terminal velocity approximation and the full one-fluid method for weakly coupled grains. However, the effect is substantially weaker in the terminal velocity approximation because the dust-fraction evolution remains diffusive for all grain sizes, suppressing unchecked growth of dust concentrations. In contrast, the full one-fluid evolution becomes increasingly advective for weakly coupled grains, allowing overdensities to persist and continue growing. The appearance of these clumps also revealed a separate limitation of the one-fluid formalism: particles whose dust fractions grow to large values can artificially displace gas and systematically inflate the dust fractions of all dust species, not only those directly associated with the clumping. While some efforts have been made to mitigate the effects of multivalued dust velocities through pressure-like regularisation, a more general framework is still lacking. Current dissipation schemes adapted from the gas also struggle to suppress numerical post-shock oscillations in multi-species dust systems, highlighting the potential need for dust-specific dissipation switches. Finally, the full one-fluid method is computationally expensive, requiring up to an order of magnitude more wall time than the terminal velocity approximation. Incorporating dust growth and fragmentation will only increase this burden, underscoring the importance of leveraging parallel architectures such as GPUs for future simulations.

\section*{Acknowledgements}

We thank the referee for their thorough and constructive report, which significantly improved the scope and clarity of this paper.
MH and CK acknowledge support from the DFG program ``Closing the Loop - Using Synthetic Observations of Simulated Star-forming Regions to Test Observational Properties" (DFG Project Number: 426714422). 
ML acknowledges funding from European Research Council (ERC) under the European Union’s Horizon 2020 research and innovation program (Grant agreement No. 101098309 - PEBBLES).
GL acknowledges funding from the European Research Council for the FP7 ERC advanced grant project ECOGAL.


\section*{Data Availability}

Select data, plotting scripts, and analysis routines used to generate the figures in this article -- including the 1D solvers used in the \textsc{dustywave} and \textsc{dustysettle} tests -- are available on Zenodo via the DOI: \href{https://doi.org/10.5281/zenodo.20525185}{10.5281/zenodo.20525185}.



\bibliographystyle{mnras}
\bibliography{bibliography} 




\appendix

\section{SPH derivation}
\label{sec:sph_derivation}

In this section, we present the full derivation of the conservative SPH relations that solve \cref{eq:genmass_rho,eq:gendtgevol,eq:genmomentum_bary,eq:genmomentum_deltav,eq:newusingle}.

\subsection{Conserved quantities}
\label{sec:conserved}

We start constructing the SPH formalism based on the conservation of physical quantities over the volume $V$ occupied by the mixture, namely: the total mass $M$, gas mass $M_{\rm g}$, dust mass of each species $\Mdj$, linear momentum $\mathbf{p}$, and energy $E$
\begingroup
\allowdisplaybreaks
\begin{align}
    M \equiv {}& \int \rho {\rm d}V = \sum_{a} m_{a}
    \label{eq:newm},
\\
    M_{\rm g}  \equiv {}& \int \rhog {\rm d}V = \int \left(1 - \epsd \right) \rho {\rm d}V = \sum_{a} m_{a} \left(1 - \epsda \right) ,
    \label{eq:mg} 
\\
    \Mdj  \equiv {}& \int \rhodj {\rm d}V = \int \epsj \rho {\rm d}V  = \sum_{a} m_{a} \epsdja
    \label{eq:mdj} .
\\
    \mathbf{p}  \equiv {}& \int \rho \vb {\rm d}V = \sum_{a} m_{a} {\bf v}_{a}
    \label{eq:newp},
\\
    E  \equiv {}& \int \frac{\rho}{2} \left[  \vb^{2} +   \sumj \epsj \deltavj^{2} - \left( \epsd \deltav \right)^{2}  + 2 \left(1 - \epsd \right) u  \right] {\rm d}V  ,
    \nonumber
\\
     = {}& \sum_{a} \frac{m_{a}}{2}  \left[ \vb_{a}^{2} +   \sumj \epsdja \deltavja^{2} - \left( \epsda \deltava \right)^{2}  +  2 \left(1 - \epsda \right) u_{a}  \right].
    \label{eq:newe}
\end{align}
\endgroup
We then proceed by systematically translating \cref{eq:genmass_rho,eq:gendtgevol,eq:genmomentum_bary,eq:genmomentum_deltav,eq:newusingle} into SPH relations, using the conservation equations above to ensure that the various terms are correctly-balanced.

\subsection{Densities}
\label{sec:densities}

The continuity equation~(\ref{eq:genmass_rho}) is in standard form and can be discretised in the usual way
\begin{equation}
    \rho_{a} = \sum_{b} m_{b} W_{ab} (h_{a}),
    \label{eq:rhosum}
\end{equation}
where $W_{ab}(h) \equiv W(\vert {\bf r}_{a} - {\bf r}_{b} \vert, h)$ is the smoothing kernel and $h_{a}$ the smoothing length. The latter is computed according to
\begin{equation}
    h_{a} = \eta \left(\frac{m_{a}}{\rho_{a}}\right)^{1/\nu},
    \label{eq:hfac}
\end{equation}
which adapts the number of SPH neighbours to the local SPH density. For simulations with multiple dust species, the resolution of the simulation will always follow the total mass of the mixture, which may be dominated by the gas mass or by one (or more) of the dust components. We designate $\nu$ as the number of spatial dimensions and set $\eta = 1.2$ to optimise the interpolation errors \citep[see][]{Price/2012}. \cref{eq:rhosum,eq:hfac} are solved simultaneously using an iterative Newton-Raphson method \citep{Price/Monaghan/2004,Price/Monaghan/2007}.

\subsection{Dust fraction equation}
\label{sec:dustfrac}

\cref{eq:newm} for the total mass of the system is trivially conserved by virtue of having constant mass particles, but the total mass of each species (Equations~\ref{eq:mg} and \ref{eq:mdj}) depend on the dust fractions $\epsdja$. Taking the derivative of \cref{eq:mdj} and setting it equal to zero gives the constraint on the dust fractions that we need to conserve the dust mass
\begin{equation}
    \frac{\mathrm{d} \Mdj}{{\rm d}t} =  \sum_{a} m_{a}  \frac{{\rm d} \epsdja }{{\rm d} t} = 0.
    \label{eq:dmdustdt}
\end{equation}
With the total particle mass and the mass of each individual dust species both conserved, it follows that the gas mass must also be conserved. We can show this explicitly by summing \cref{eq:dmdustdt} over the index $j$
\begin{equation}
    \sum_{a} m_{a} \sumj  \frac{{\rm d} \epsdja }{{\rm d} t} = \sum_{a} m_{a} \frac{{\rm d} \epsda }{{\rm d} t} = \frac{\mathrm{d} \Mg}{{\rm d}t} =  0 .
    \label{eq:dmgasdt}
\end{equation}
We can enforce these conditions in SPH by employing an antisymmetric argument in $a$ and $b$ for $ \frac{{\rm d} \epsdja }{{\rm d} t}$ (i.e. by associating a symmetric argument for the SPH divergence and an antisymmetric kernel gradient), since the double summation of an antisymmetric expression is zero. Thus,
\begin{align}
    \frac{{\rm d}\epsdja}{{\rm d} t} = - \sum_{b} m_{b} & \left[ \frac{\epsdja }{\Omega_{a} \rho_{a}} \left(\deltavja - \epsda \deltava \right)\cdot\nabla_{a} W_{ab} (h_{a})\right.
    \nonumber
\\
    & + \left.  \frac{\epsdjb }{\Omega_{b} \rho_{b}}\left(\deltavjb - \epsdb \deltavb \right)\cdot \nabla_{a} W_{ab} (h_{b}) \right], 
    \label{eq:dusttogassph}
\end{align}
where $\Omega_{a}$ is the usual term that accounts for the variable smoothing length in kernel derivatives,
\begin{equation}
    \Omega_{a} = 1 - \frac{\partial h_{a}}{\partial\rho_{a}} \sum_{b} m_{b} \frac{\partial W_{ab}(h_{a})}{h_{a}}.
\end{equation}
\cref{eq:dusttogassph} is the SPH equivalent of \cref{eq:gendtgevol}. Although we do not evolve the total dust fraction directly, summing over $j$ in \cref{eq:dusttogassph} gives
\begin{align}
    \frac{{\rm d}\epsda}{{\rm d} t} = - \sum_{b} m_{b} & \left[ \frac{\left( 1 - \epsd_{a} \right)}{\Omega_{a} \rho_{a}} \epsda \deltava\cdot\nabla_{a} W_{ab} (h_{a})\right.
    \nonumber
\\
    & + \left.   \frac{\left( 1 - \epsd_{b} \right)}{\Omega_{b} \rho_{b}} \epsdb \deltavb\cdot\nabla_{a} W_{ab} (h_{b}) \right], 
    \label{eq:dusttogassph_full}
\end{align}
where we have used the identity
\begin{equation}
    \sumj \epsd_{j} \left( \deltavj - \sumk \deltavk \epsd_{k} \right) = \left(1 - \epsd \right) \epsd \deltav .
    \label{eq:gasident}
\end{equation}
%

\subsection{Momentum equation}
\label{sec:momentum}

At this point, we will proceed assuming that there are no dust forces (i.e. $\fdsum = 0$) and that the only gas forces are the usual hydrodynamic forces, i.e.
\begin{equation}
    \fg = - \frac{\nabla P}{\rhog} + \fgvisc,
\end{equation}
where $\fgvisc$ is a viscosity term whose SPH representation will hereafter be written alongside the pressure. As we did with the mass before, we take the derivative of \cref{eq:newp} and set it equal to zero in order to obtain a constraint on the momentum equation. The SPH translation of \cref{eq:genmomentum_bary} is constructed such that a double summation over the gas pressure gradient (using the usual SPH formalism) and anisotropic pressure gradient terms (specific to dust and gas mixtures) gives exactly zero, namely
\begin{align}
    \frac{{\rm d} \vb_{a}}{{\rm d} t} = & -\sum_{b} m_{b} \left[ \frac{P_{a} + q^{\rm AV}_{v, a}}{\Omega_{a} \rho_{a}^{2}} \nabla_{a} W_{ab}(h_{a}) + \frac{P_{b} + q^{\rm AV}_{v, b}}{\Omega_{b} \rho_{b}^{2}} \nabla_{a} W_{ab}(h_{b}) \right] 
    \nonumber
\\
    & -\sum_{b} m_{b} \left[ \sumj \frac{ \epsdja \deltavja}{\Omega_{a} \rho_{a}} \left(\deltavja - \epsda \deltava \right)\cdot\nabla_{a} W_{ab}(h_{a}) \right.
    \nonumber
\\
    &  \phantom{\sum_{b} m_{b}[} + \sumj \left.\frac{\epsdjb\deltavjb}{\Omega_{b} \rho_{b}} \left(\deltavjb - \epsdb \deltavb \right)\cdot\nabla_{a} W_{ab}(h_{b}) \right]  + {\bf f}_{a},
    \label{eq:sphmom}
\end{align}
where the $q^{\rm AV}_{v, a}$ and $q^{\rm AV}_{v, b}$ terms represent the artificial viscosity (discussed in \cref{sec:artificial_viscosity}, below).

\subsection{Differential velocity and energy equations}
\label{sec:differential velocities}

The SPH expressions for \cref{eq:genmomentum_deltav,eq:newusingle} come directly from the total energy conservation, i.e. $\frac{\mathrm{d}E_{a}}{\mathrm{d}t} = 0$. From \cref{eq:newe}, this condition reduces to
\begin{align}
    \sum_{a} m_{a} \Bigg[ &{\bf v}_{a}. \frac{{\rm d}{\bf v}_{a}}{{\rm d}t} +
    \sumj \epsdja \left(\deltavja - \epsda \deltava \right) \cdot \frac{{\rm d}{\deltavja}}{{\rm d}t}
    \nonumber
\\
    & +\sumj \frac{\deltavja}{2}\cdot \left( \deltavja - 2 \epsda \deltava \right) \frac{{\rm d}\epsdja}{{\rm d}t}
    \nonumber
\\
    &  - u_{a}\frac{{\rm d}\epsd_{a}}{{\rm d}t}
    + \left(1 - \epsda \right) \frac{{\rm d}u_{a}}{{\rm d}t}
    \Bigg] = 0.
    \label{eq:dedt}
\end{align}
From here, it is easier to separately match and cancel the contributions from corresponding terms in the different fluid equations.

\subsubsection{Pressure gradient: $\fg$ and $P{\rm d}V$ work}
\label{sec:fgas_pdv_work}

The contributions from the pressure force in the $\frac{{\rm d}{\bf v}}{{\rm d}t}$ and $\frac{{\rm d}\deltavj}{{\rm d}t}$ equations on the gas combine to generate the $P{\rm d}V$ work term in the energy equation (Equation~\ref{eq:newusingle}).
\begin{align}
    \sum_{a} m_{a} & \left(1 - \epsda \right) \left( \frac{{\rm d}u_{a}}{{\rm d}t}\right)_{P{\rm d}V} 
    \nonumber
\\
    &=  - \sum_{a} m_{a} \left[\left(1 - \epsda \right){\bf v}_{a} - \sumj \epsdja \left( \deltavja - \epsda \deltava \right)\right] \cdot \fg ,
    \nonumber
\\
    & =  - \sum_{a} m_{a} \left(1 - \epsda \right){\bf v}_{{\rm g},a}\cdot \fg ,
    \label{eq:pdvterm}
\end{align}
where in the second equality we used \cref{eq:gasident,eq:vgas} to simplify the term in square brackets. Importantly, the expression obtained in \cref{eq:pdvterm} is rigorously identical to the single-dust-species case. We thus follow \citet{Laibe/Price/2014b} to get the expression of the $\left(\frac{{\rm d}u_{a}}{{\rm d}t}\right)_{P{\rm d}V}$ term. In short, we substitute in the expression for $(1-\epsd)\fg$ (i.e. the first summation in Equation~\ref{eq:sphmom}), swap summation indices in the resulting double summation from the second term, and use the antisymmetry of the kernel gradient ($\nabla_{a} W_{ab} = - \nabla_{b} W_{ba}$) to obtain
\begin{equation}
    \left(\frac{{\rm d}u_{a}}{{\rm d}t}\right)_{P{\rm d}V} = \frac{P_{a} + q^{\rm AV}_{v, a}}{\Omega_{a} \rho_{a} \rho_{{\rm g},a}} \sum_{b} m_{b} \left({\bf v}_{{\rm g},a} - {\bf v}_{{\rm g},b} \right) \cdot \nabla_{a} W_{ab}(h_{a}).
\end{equation}
This term is the total energy conserving SPH equivalent of the pressure gradient term in \cref{eq:newusingle}.

\subsubsection{Anisotropic pressure: $-(\Delta{\bf v}_{j} \cdot\nabla){\bf v}$}
\label{sec:deltav_term}

The work performed by the anisotropic pressure contributes to an equivalent $P{\rm d}V$ energy term. Physically, this comes from the difference between the kinetic energy of the two phases versus the kinetic energy of the mixture, which, in addition to the regular pressure term, contributes to the total internal energy of the mixture in the barycentric reference frame. Balancing the first and second terms of \cref{eq:dedt} provides
\begin{align}
    &\sum_{a} m_{a}  \sumj \epsdja   \left(\deltavja -  \epsda \deltava \right) \cdot \left( \frac{{\rm d}{\deltav}_{a}}{{\rm d}t} \right)_{\mathrm{aniso}} 
    \nonumber
\\
    &  = -\sum_{a} m_{a} {\bf v}_{a} \cdot \left( \frac{{\rm d}{\bf v}_{a}}{{\rm d}t} \right)_{\mathrm{aniso}} , 
    \nonumber
\\
    &=\sum_{a} m_{a} {\bf v}_{a}\cdot \sum_{b} m_{b} \sumj \left[  \frac{\epsdja  \left(\deltavja -  \epsda \deltava \right) }{\Omega_{a} \rho_{a}} \deltavja  \cdot \nabla_{a} W_{ab}(h_{a}) \right.
    \nonumber
\\
    & \phantom{+  \sum_{a} m_{a} {\bf v}_{a}\cdot \sum_{b} } 
    \left. + \frac{\epsdjb \left(\deltavjb - \epsdb \deltavb \right) }{\Omega_{b} \rho_{b}} \deltavjb \cdot \nabla_{a} W_{ab}(h_{b}) \right], 
    \label{eq:balance_aniso}
\end{align}
the second equality coming from substituting the corresponding anisotropic contribution from $\frac{{\rm d}{\bf v}_{a}}{{\rm d}t}$ (i.e. the second summation in Equation~\ref{eq:sphmom}).
As before, we swap the summation indices in the second term, use the antisymmetry of the kernel gradient, and rearrange. We can then read off the energy-conserving relation for the anisotropic terms
\begin{equation}
    -(\deltavj\cdot\nabla){\bf v} \approx \frac{1}{\rho_{a}\Omega_{a}}\sum_{b} m_{b} ({\bf v}_{a} - {\bf v}_{b})  \deltavja \cdot \nabla_{a} W_{ab} (h_{a}).
    \label{eq:back_adv}
\end{equation}
%

\subsubsection{Dust fraction: $(\Delta{\bf v} \cdot \nabla)u$ and $\nabla \left[ \Delta{\bf v}_j \cdot \left(\Delta{\bf v}_j - 2 \epsd \Delta{\bf v} \right) \right]$}
\label{sec:more_deltav_terms}

The last two contributions to the energy conservation (Equation~\ref{eq:dedt}) come from the contributions of the dust fractions and the differential velocity. Focusing on the dust fractions first,
\begin{equation}
    \sum_{a} m_{a}\left(1 - \epsda \right) \left( \frac{{\rm d}u_{a}}{{\rm d}t}\right)_{\epsd} = \sum_{a} m_{a} u_{a} \frac{{\rm d}\epsd_{a}}{{\rm d}t}.
\end{equation}
Substituting from \cref{eq:dusttogassph_full} and rearranging the double summation as we have hitherto done, we obtain
\begin{equation}
    \left( \frac{{\rm d}u_{a}}{{\rm d}t}\right)_{\epsd} = - \frac{1}{\Omega_{a} \rho_{a}}  \sum_{b} m_{b} (u_{a} - u_{b}) \epsda \deltava \cdot \nabla_{a} W_{ab} (h_{a}) ,
\end{equation}
which is the SPH representation of $\epsd \deltav \cdot \nabla u$ from \cref{eq:newusingle}.

As for the differential velocity, we balance the second and third terms of \cref{eq:dedt} to give
\begin{align}
    \sum_{a} m_{a} &  \sumj \epsdja \left(\deltavja - \epsda \deltava \right)\cdot \left(\frac{{\rm d}\deltavja}{{\rm d}t} \right)_{\epsd} =
    \nonumber
\\
    -\sum_{a} m_{a} & \sumj \frac{\deltavja}{2} \left(\deltavja - 2 \epsda \deltava \right)  \frac{{\rm d}\epsdja}{{\rm d}t}.
    \label{eq:deltav_dustfrac_conjugate}
\end{align}
Substituting from \cref{eq:dusttogassph} and rearranging the double summation provides
\begin{align}
    \left(\frac{{\rm d}\deltavja}{{\rm d}t} \right)_{\epsd} = {}& \frac{1}{2\rho_{a}\Omega_{a}}  \sum_{b} m_{b}  \left[ \deltavja \cdot \left(\deltavja - 2 \epsda \deltava \right) \right. 
    \nonumber
\\
    & \left. -  \deltavjb \cdot \left(\deltavjb - 2 \epsdb \deltavb \right)  \right] \nabla_{a} W_{ab} (h_{a}),
    \label{eq:deltav_dustfrac_contribution}
\end{align}
which is the SPH equivalent of $-\frac{1}{2}\nabla \left[\deltavj \cdot \left(\deltavj - 2 \epsd \deltav \right) \right]$.

\subsubsection{Drag terms}
\label{sec:drag_terms}

The last two terms in \cref{eq:genmomentum_deltav} are related to the drag and account for the dissipation of differential velocity into heat. Drag contributions to the internal energy of the gas are often negligible in astrophysical environments; nevertheless, for strict energy conservation, \cref{eq:dedt} implies
\begin{align}
    \sum_{a} m_{a} & \left(1 - \epsda \right) \left( \frac{{\rm d}u_{a}}{{\rm d}t}\right)_{\rm drag} = 
    \nonumber
\\    
    &- \sum_{a} m_{a} \sumj \left[ \epsdja \left(\deltavja - \epsda \deltava \right) \cdot \left( \frac{\mathrm{d}\deltavja}{\mathrm{d} t} \right)_{\rm drag} \right].
\end{align}
Inserting the straightforward SPH translations of the drag terms in \cref{eq:genmomentum_deltav} and dividing through by $1-\epsda$, gives the SPH relation for the drag heating in the gas
\begin{align}
    \left( \frac{{\rm d}u_{a}}{{\rm d}t}\right)_{\rm drag} & =  \frac{1}{\left(1 - \epsda \right)} \sumj \Bigg[  \epsdja \left(\deltavja - \epsda \deltava \right) 
    \nonumber
\\
     & \phantom{\frac{1}{\left(1 - \epsda \right)} \sumj \Bigg[}
     \cdot \left(\frac{\deltavja}{ \epsdja \tja}  + \sumk \frac{\deltavka}{\left(1 - \epsda \right) \tka}\right) \Bigg] ,
     \nonumber
\\
    &=  \sumj \frac{\deltavja^{2}}{\left(1 - \epsda \right)\tja},
    \label{eq:du_drag}
\end{align}
where the second equality comes from applying \cref{eq:def_epsilon,eq:def_deltav} and simplifying.

\subsubsection{Remaining terms}
\label{sec:remaining_terms}

At this point, we still have four continuum terms in \cref{eq:genmomentum_deltav} that are unaccounted for in our SPH relations, but we have exhausted the available conservation relations from \cref{sec:conserved}. This suggests that the energy contributions from the remaining terms must cancel among themselves when substituted into \cref{eq:dedt}. More precisely, the remaining terms must satisfy the following relation
\begin{equation} 
    \sumj \epsdja \left(\deltavja - \epsda \deltava \right) \cdot \left(\frac{{\rm d}{\deltavja}}{{\rm d}t}\right)_{\rm null} = 0.
    \label{eq:energy_terms_constraint}
\end{equation}
Beyond this criterion, this is no restriction on how we construct the SPH relations. 
Using the same formalism we used for \cref{eq:deltav_dustfrac_contribution}, the SPH relations for the final four terms can be written as
\begin{align}
    \left(\frac{{\rm d}{\deltavja}}{{\rm d}t}\right)_{\rm null} &=  \phantom{-} \frac{1}{\rho_{a} \Omega_{a}} \sum_{b} m_{b} \epsda \deltava \cdot (\deltavja - \deltavjb) \nabla_{a} W_{ab} (h_{a}) 
    \nonumber
\\
    {} & \phantom{=} - \frac{1}{\rho_{a} \Omega_{a}} \sum_{b} m_{b} \deltavja \cdot [(\deltavja - \epsda \deltava) 
    \nonumber
\\
    {} & \phantom{= - \frac{1}{\rho_{a} \Omega_{a}} \sum_{b} }
    - (\deltavjb - \epsdb \deltavb)] \nabla_{a} W_{ab} (h_{a}) 
    \nonumber
\\
    {} &\phantom{=}- \frac{1}{\rho_{a} \Omega_{a}} \sum_{b} m_{b} (\deltavja - \deltavjb) \epsda \deltava \cdot \nabla_{a} W_{ab} (h_{a}) 
    \nonumber
\\
    {} & \phantom{=}+ \frac{1}{\rho_{a} \Omega_{a}} \sum_{b} m_{b} [(\deltavja - \epsda \deltava)
    \nonumber
\\
    {} & \phantom{=- \frac{1}{\rho_{a} \Omega_{a}} \sum_{b}}
    - (\deltavjb - \epsdb \deltavb)] \deltavja \cdot  \nabla_{a} W_{ab} (h_{a}) .
    \label{eq:deltav_sph_null}
\end{align}
Inserting the above terms into \cref{eq:energy_terms_constraint}, it is straightforward to show that all terms indeed cancel. A similar exercise shows that the criterion holds for the continuum relations as well. 

It is worth mentioning that the evolution equation for the differential velocity (Equation~\ref{eq:genmomentum_deltav}) can be rewritten in a plethora of ways, many of which appear significantly simpler in form. Unfortunately, the terms in these simpler expressions no longer exhibit the same conjugate pairing we exploited in \cref{eq:deltav_dustfrac_conjugate}. While it may be possible to rewrite \cref{eq:genmass_rho,eq:gendtgevol,eq:genmomentum_bary,eq:genmomentum_deltav,eq:newusingle} in a way that enforces conservation for these simpler forms, there is at least one major advantage of using the equations we have presented. They closely resemble the equations used in the implementation of the multi-species, one-fluid equations under the terminal velocity approximation in \citet{Hutchison/Price/Laibe/2018}, making it easier to code and maintain both methods.

\subsection{Shock-capturing terms}
\label{sec:shock-capturing_terms}

\subsubsection{Artificial viscosity}
\label{sec:artificial_viscosity}

The artificial viscosity term is computed as follows
\begin{equation}
    q^{\rm AV}_{v, a} =
	\begin{cases}
		-\frac{1}{2} \left( 1 - \epsilon_{a} \right) v^{\mathrm{sig}}_{v,a} {\bf v}_{ab}^{\rm g
        } 
			\cdot \hat{{\bf r}}_{ab}, & \qquad {\bf v}_{ab}^{\rm g} \cdot \hat{{\bf r}}_{ab} < 0
	\\	
		0,  & \qquad \mathrm{otherwise},
	\end{cases}
\end{equation}
where ${\bf v}_{ab}^{\rm g} \equiv {\bf v}_{a}^{\rm g} - {\bf v}_{b}^{\rm g}$ (similarly for $\hat{{\bf r}}_{ab}$) and the signal speed $v^{\mathrm{sig}}_{v}$ corresponds to the usual choice for hydrodynamics, i.e.
\begin{equation}
    v^{\mathrm{sig}}_{v,a} = \alpha_{v,a} c_{\mathrm{s},a} + \beta_{v} \vert {\bf v}_{ab}^{\rm g} \cdot \hat{{\bf r}}_{ab} \vert,
\end{equation}
where $\alpha_{v,a} \in [0,1]$ is the linear dimensionless viscosity parameter \citep[the particle index $a$ implying that $\alpha_{v}$ can be unique to each particle; see, e.g.,][]{Morris/Monaghan/1997,Cullen/Dehnen/2010}, $c_{\rm s}$ is the gas sound speed, and $\beta^{\rm AV}$ (typically $\beta^{\rm AV} = 2$) is the von Neumann-Richtmyer viscosity parameter.

\subsubsection{Artificial conductivity}
\label{sec:artificial_conductivity}

In order to correctly treat contact discontinuities, an artificial conductivity term must be added to the energy equations \citep[see][]{Price/2008},
\begin{equation}
    \left( \frac{\mathrm{d} u_{a}}{\mathrm{d} t} \right)_{\rm cond} = \frac{1}{1-\epsilon_{a}} \sum_{b} m_{b}
		\left[ \frac{Q^{\rm AC}_{u, a}}{\Omega_{a} \rho_{a}^{2}} F_{ab} (h_{a}) + 
		\frac{Q^{\rm AC}_{u, b}}{\Omega_{b} \rho_{b}^{2}} F_{ab} (h_{b}) \right],
\end{equation}
where
\begin{equation}
    Q^{\rm AC}_{u,a} = \frac{1}{2} \alpha_u \rho_{a} v^{\mathrm{sig}}_{u} \left( u_{a} - u_{b} \right),
\end{equation}
with $\alpha_u \in [0,1]$ the dimensionless conductivity parameter and a signal speed given by \citep[see][]{Price/2008,Wadsley/Veeravalli/Couchman/2008}
\begin{equation}
    v^{\mathrm{sig}}_{u} = \alpha_{u}
    \begin{cases}
        \sqrt{\frac{|P_{a} - P_{b}|}{\overline{\rho}_{ab}}}, & \qquad {\rm without\;gravity}
    \\
        |{\bf v}_{ab} \cdot \hat{{\bf r}}_{ab}|, & \qquad {\rm with \; gravity.}
    \end{cases}
\end{equation}
%

\subsubsection{Artificial dissipation}
\label{sec:artificial_dissipation}

\citet{Laibe/Price/2014b} noted that post-shock oscillations in the differential velocity would persist in the weak-coupling limit (i.e. large $\deltav$) without some sort of artificial dissipation. A formulation that conserves both energy and momentum can be constructed as follows
\begin{equation}
    \left( \frac{\mathrm{d} \deltavja}{\mathrm{d} t} \right)_{\rm diss} = \sum_b  m_b   \left[ \frac{q^{\rm AD}_{\Delta v,ja}}{\Omega_{a} \rho_{a}^{2}}  \nabla_{a} W_{ab}(h_{a}) 		+ \frac{q^{\rm AD}_{\Delta v,jb}}{\Omega_{b} \rho_{b}^{2}}  \nabla_{a} W_{ab}(h_{b})  \right] ,
\end{equation}
with the accompanying heating term in the energy being
\begin{align}
    \left( \frac{\mathrm{d} u_{a}}{\mathrm{d} t} \right)_{\rm diss} = & - \frac{1}{1-\epsda}\sum_{b}  m_{b}   \frac{q^{\rm AD}_{\Delta v,ja}}{\Omega_{a} \rho_{a}^{2}} \Big[ \epsdja (\deltavja - \epsda \deltava)
    \nonumber
\\    
    & \phantom{- \frac{1}{1-\epsda}} - \epsdjb (\deltavjb - \epsdb \deltavb)\Big] \cdot \mathbf{\hat{r}}_{ab}  F_{ab}(h_{a}),
\end{align}
the dissipation parameter being
\begin{equation}
    q^{\rm AD}_{\Delta v,ja} = \frac{1}{2} \rho_a v^{\rm sig}_{\Delta v, a} \Big[ \epsdja (\deltavja - \epsda \deltava) - \epsdjb (\deltavjb - \epsdb \deltavb)\Big] \cdot \mathbf{\hat{r}}_{ab},
\end{equation}
with a signal velocity
\begin{equation}
    v^{\rm sig}_{\Delta v, a} = \alpha_{\Delta v} c_{{\rm s},a},
\end{equation}
where $ \alpha_{\Delta v}$ is a dimensionless coefficient of order unity.

\section{Numerical benchmarking tool for the \textsc{dustywave} Test}
\label{sec:dustywave_solution}

The dispersion relation for the \textsc{dustywave} problem with multiple dust species is given in Section 3.2 of \citet{Benitez-Llambay/Krapp/Pessah/2019}, which, along with the system's eigenvectors, provides a full analytic solution. However, applying this solution to arbitrary setups is tedious because it requires expressing the initial conditions as a superposition of eigenmodes. Moreover, since we plan to incorporate grain-size evolution into our dust models in the near future and will need a numerical benchmarking tool at that stage anyway (due to the time-dependent dust mass distribution), we decided to develop a lightweight 1D numerical solver for the \textsc{dustywave} test in advance.

\subsection{Spectral Method}
\label{sec:spectral_method}

Finding a numerical solution is made significantly easier if we first linearise  \cref{eq:genmass_rho,eq:gendtgevol,eq:genmomentum_bary,eq:genmomentum_deltav} about the equilibrium solution $\rho = \rho_0$, $\epsd=\epsd_0$, and $v=\Delta v=0$. Following \citet{Laibe/Price/2014c} and using an isothermal equation of state $\delta P = \cs^2[(1-\epsd_0)\delta \rho-\rho_0\delta \epsd]$, we get
\begin{align}
    \frac{\partial \delta \rho}{\partial t} &= -\rho_0 \frac{\partial \delta v}{\partial x},
    \label{eq:dustywave_lin_mass} 
\\
    \frac{\partial \delta \epsilon_j}{\partial t} &= -\epsilon_{j0} \frac{\partial \delta \Delta v_j}{\partial x} + \epsilon_{j0} \sum_k \epsilon_{k0} \frac{\partial \delta \Delta v_k}{\partial x},
    \label{eq:dustywave_lin_dustfrac}
\\
    \frac{\partial \delta v}{\partial t} &= -\frac{c_s^2 (1 - \epsilon_0)}{\rho_0} \frac{\partial \delta \rho}{\partial x} + c_s^2 \frac{\partial \delta \epsilon}{\partial x},
    \label{eq:dustywave_lin_momentum} 
\\
    \frac{\partial \delta \Delta v_j}{\partial t} &= -\frac{\delta \Delta v_j}{t_j \epsilon_{j0}} - \frac{1}{1 - \epsilon_0} \sum_k \frac{\delta \Delta v_k}{t_k} + \frac{c_s^2}{\rho_0} \frac{\partial \delta \rho}{\partial x} - \frac{c_s^2}{1 - \epsilon_0} \frac{\partial \delta \epsilon}{\partial x},
    \label{eq:dustywave_lin_deltav} 
\end{align}
where $\delta$ is used to identify quantities that are first-order small. This system of equations can be written more compactly by introducing the vector ${\mathbf u}(x, t)$
\begin{align}  
    {\mathbf u}(x, t) =& (\delta \rho(x, t), \delta v(x, t), \delta \epsilon_1(x, t), \dots, \delta \epsilon_N(x, t), 
    \nonumber
\\
    & \phantom{(} \delta \Delta v_1(x, t), \dots, \delta \Delta v_N(x, t))^\top,  
\end{align}  
defined at time $t$ and coordinate $x$ for $N$ dust species. The system of equations can be rewritten as a vectorial partial differential equation by introducing the linear operator $\mathcal{D}_N$, which transforms the system into the form  
\begin{equation}  
    \frac{\partial {\mathbf u}}{\partial t} = \mathcal{D}_N[t, {\mathbf u}, \partial_x {\mathbf u}] = A_N(t) \frac{\partial {\mathbf u}}{\partial x} + B_N(t) {\mathbf u},  
    \label{eq:vectorial_pde}
\end{equation}  
where $A_N(t)$ and $B_N(t)$ are matrices composed of coefficients from the original equations and can, in general, be time dependent if any of the coefficients depend on time. Details on how to construct these coefficient matrices are given in \cref{sec:matrices}.   

A computationally efficient approach to solving \cref{eq:vectorial_pde} is to apply a spectral method, which takes advantage of the periodicity of the solutions. By expressing all perturbations in Fourier space,
\begin{equation}
    \hat{{\mathbf u}}(k,t) = \int {\mathbf u}(x,t) e^{-i k x} \,dx,
\end{equation}
where $k$ in this context is the wave number, the system transforms from a set of partial differential equations into a set of ordinary differential equations in time. \Cref{eq:vectorial_pde} can be rewritten as  
\begin{equation}
    \frac{{\rm d}\hat{{\mathbf u}}(k,t)}{{\rm d}t} = \hat{\mathcal{D}}_N(k,t) \hat{{\mathbf u}}(k,t),
    \label{eq:ode_single_k}
\end{equation}
and the linear operator in Fourier space, $\hat{\mathcal{D}}_N(k,t)$, reduces to a constant matrix
\begin{equation}
    \hat{\mathcal{D}}_N(k,t) = i k A_N(t) + B_N(t).
\end{equation}

In practice, we solve an expanded version of \cref{eq:ode_single_k} for all wavenumbers at once by introducing the following matrix with dimensions $[N_k,\, 2N+2]$, 
\begin{equation}  
    W_{\kappa\alpha}^n \equiv \hat{u}_\alpha(k_\kappa, t_n),  
\end{equation}
which stores the Fourier-transformed solution for all discrete wavenumbers $k_\kappa$ at time step $t_n$. Here, $\kappa \in [1,N_k]$ indexes the discrete wavenumbers, while $\alpha$ (and later $\beta$) runs over the $2N+2$ components of the state vector (i.e. the perturbed gas density, velocity, dust fractions, and differential velocities). Similarly expanding our linear Fourier-space operator into a tensor with dimensions $[N_k,\,2N+2,\,2N+2]$, 
\begin{equation}  
    \hat{\mathcal{D}}_{\kappa\alpha\beta}(t) \equiv \left[ \hat{\mathcal{D}}_N(k_\kappa, t) \right]_{\alpha\beta},  
\end{equation}  
we can rewrite \cref{eq:ode_single_k} as  
\begin{equation}  
    \frac{{\rm d} W_{\kappa\alpha}^n}{{\rm d}t} = \sum_{\beta=1}^{2N+2} \hat{\mathcal{D}}_{\kappa\alpha\beta}(t) W_{\kappa\beta}^n,  
    \label{eq:ode_all_k}
\end{equation}  
which we evolve in time using an explicit fourth-order Runge-Kutta solver. To efficiently compute the right-hand side of \cref{eq:ode_all_k}, we use NumPy's \texttt{einsum} function, which optimizes the summation operation in a vectorized manner, reducing computational overhead. As a final detail, we transition between real and Fourier space by applying the Fast Fourier Transform (FFT) to the real-space initial conditions at $t=0$ and its inverse (iFFT) to $k$-space solutions $W^n$ at $t_n$.

\subsection{Coefficient Matrices}
\label{sec:matrices}

Here we define the coefficient matrices introduced in the previous section. Let the identity matrix be denoted by $I_N \in \mathcal{M}_{N,N}$, where $\mathcal{M}_{m,n}$ is the space of $m \times n$ matrices. Additionally, we define a ones matrix $1_N \in \mathcal{M}_{N,N}$, which consists entirely of ones, and the zero matrix $\mathbb{O}_{l,m} \in \mathcal{M}_{l,m}$ consisting entirely of zeros, with the special case $\mathbb{O}_N = \mathbb{O}_{N,N}$. The matrix $A_N$ can be then be written in the succinct form
\begin{equation}
    A_N =
    \begin{bmatrix}
        M^{(1)} & M^{(2)} & \mathbb{O}_{N,2}
    \\
        \mathbb{O}_{2,N} & \mathbb{O}_N & D_N
    \\
        M^{(3)} & E_N & \mathbb{O}_N
    \end{bmatrix}
    \in \mathcal{M}_{2N+2, 2N+2},
\end{equation}
where the remaining block matrices are defined as follows
\begin{align}
    M^{(1)} =& 
    \begin{bmatrix} 
        0 & -\rho_0
    \\
        - \frac{c_s^2 (1 - \epsilon_0)}{\rho_0} & 0 
    \end{bmatrix},
\\
    M^{(2)} =& 
    \begin{bmatrix} 
        0 & \cdots & 0
    \\ 
        c_s^2 & \cdots & c_s^2
    \end{bmatrix},
\\
    M^{(3)} =& 
    \begin{bmatrix} 
        \frac{c_s^2}{\rho_0} & 0 
    \\ 
        \vdots & \vdots 
    \\ 
        \frac{c_s^2}{\rho_0} & 0 
    \end{bmatrix}.
\\
    D_N =& -\text{Diag}[(\epsilon_{j0})_j] \left[ I_N - \begin{bmatrix}
        \epsilon_{10} & \cdots & \epsilon_{N0}
    \\  
        \vdots & \ddots & \vdots
    \\
        \epsilon_{10} & \cdots & \epsilon_{N0} 
    \end{bmatrix} \right],
\\
    E_N =& - \frac{c_s^2}{1 - \epsilon_0} 1_N.
\end{align}
The matrix $B_N$ can be expressed similarly
\begin{equation}
    B_N = 
    \begin{bmatrix} 
        \mathbb{O}_{2+N} & \mathbb{O}_{N,2+N} 
    \\ 
        \mathbb{O}_{2+N,N} & F_N 
    \end{bmatrix},
\end{equation}
where the matrix $F_N$ takes the form
\begin{equation}
    F_N = -\text{Diag} \left[ \left( \frac{1}{t_j \epsilon_{j0}} \right)_j \right] - \frac{1}{1 - \epsilon_0} 
    \begin{bmatrix} 
        1/t_1 & \cdots & 1/t_N 
    \\ 
        \vdots & \ddots & \vdots 
    \\ 1/t_1 & \cdots & 1/t_N 
    \end{bmatrix}.
\end{equation}
%


\bsp	
\label{lastpage}
\end{document}